\documentclass[12pt,notoc]{JHEP3}

\usepackage{amsmath,amssymb,euscript,array,cite}

\input{epsf}
\newcommand{\startappendix}{
\setcounter{section}{0}
\renewcommand{\thesection}{\Alph{section}}}
\newcommand{\Appendix}[1]{
\refstepcounter{section}
\begin{flushleft}
{\large\bf Appendix \thesection: #1}
\end{flushleft}}
\usepackage{epsfig}

\def\c{\gamma}

\def\l{\lambda}
\def\m{\mu}
\def\n{\nu}

\def\r{\rho}
\def\s{\sigma}
\def\t{\tau}

\def\w{\omega}

\def\L{\Lambda}

\def\det{{\rm det}}

\def\Dbarslash{\,\,{\raise.15ex\hbox{/}\mkern-12mu {\bar D}}}
\def\Dslash{\,\,{\raise.15ex\hbox{/}\mkern-12mu D}}
\def\delslash{\,\,{\raise.15ex\hbox{/}\mkern-9mu \partial}}
\def\delbarslash{\,\,{\raise.15ex\hbox{/}\mkern-9mu {\bar\partial}}}

\newcommand{\EQ}[1]{\begin{equation} #1 \end{equation}}

\newcommand{\SP}[1]{\begin{equation}\begin{split} #1
\end{split}\end{equation}}

\title{The Refractive Index of Curved Spacetime: the Fate of Causality in QED}
\author{Timothy J. Hollowood and Graham M. Shore\\
Department of Physics,\\ University of Wales Swansea,\\
Swansea, SA2 8PP, UK.\\
E-mail: {\tt t.hollowood@swansea.ac.uk, g.m.shore@swansea.ac.uk}}
\abstract{
It has been known for a long time that vacuum polarization in QED leads to 
a superluminal low-frequency phase velocity for light propagating in
curved spacetime. Assuming the validity of the Kramers-Kronig dispersion
relation, this would imply a superluminal wavefront velocity and the
violation of causality. Here, we calculate for the first time the full
frequency dependence of the refractive index using world-line sigma
model techniques together with the Penrose plane wave limit of spacetime
in the neighbourhood of a null geodesic. We find that the high-frequency
limit of the phase velocity ({\it i.e.\/}~the wavefront velocity) is always
equal to $c$ and causality is assured. However, the Kramers-Kronig dispersion
relation is violated due to a non-analyticity of the refractive index in
the upper-half complex plane, whose origin may be traced to the generic 
focusing property of null geodesic congruences and the existence of
conjugate points. This puts into question the issue of  
micro-causality, {\it i.e.\/}~the vanishing of commutators of field
operators at spacelike separated points, in local quantum field theory
in curved spacetime.   
}

\begin{document}

\section{Introduction}

Quantum field theory in curved spacetime is by now a well-understood subject.
However, there remain a number of intriguing puzzles which hint at deeper 
conceptual implications for quantum gravity itself. The best known is of course
Hawking radiation and the issue of entropy and holography in quantum black 
hole physics. A less well-known effect is the discovery by Drummond and 
Hathrell \cite{Drummond:1979pp}
that vacuum polarization in QED can induce a superluminal phase velocity
for photons propagating in a non-dynamical, curved spacetime. The essential
idea is illustrated in Figure~1. Due to vacuum polarization, the photon may be
pictured as an electron-positron pair, characterized by a length scale
$\l_c = m^{-1}$, the Compton wavelength of the electron. When the curvature
scale becomes comparable to $\l_c$, the photon dispersion relation is modified.
The remarkable feature, however, is that this modification can induce a 
superluminal\footnote{In this paper, we use the term ``superluminal'' in 
the sense ``greater than $c$''.  Apart from the occasional use of $c$ in the 
text for clarity, we set $c=1$ throughout. Also, in our conventions, the metric
of flat space is $\eta=\text{diag}\,(1,-1,-1,-1)$ and the Riemann tensor is 
$R^\mu{}_{\nu\sigma\lambda}=\partial_\sigma\Gamma^\mu_{\lambda\nu}+\cdots$.} 
low-frequency phase velocity, {\it i.e.\/}~the photon momentum becomes 
spacelike.

\begin{figure}[ht] 
\centerline{\includegraphics[width=2.5in]{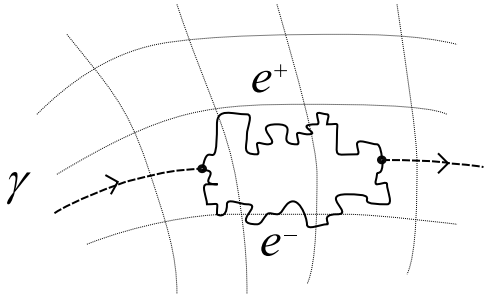}}
\caption{\footnotesize Photons propagating in curved spacetime feel the
  curvature in the neighbourhood of their geodesic because they can
  become virtual $e^+e^-$ pairs.}\label{pic5}
\end{figure}

At first, it appears that this must be incompatible with causality. However, 
as discussed in refs.~\cite{Shore:2003jx,Shore:2003zc,Shore:2007um},
the relation of causality with the ``speed of light'' is far more subtle.
For our purposes, we may provisionally consider causality to be the 
requirement that no signal may travel faster than the fundamental 
constant $c$ defining local Lorentz invariance.
More precisely, we require that the {\it wavefront velocity\/} $v_{\rm wf}$, 
defined as the speed of propagation of a sharp-fronted wave pulse, should be 
less than, or equal to, $c$. Importantly, it may be shown 
\cite{Leontovich,Shore:2003jx,Shore:2007um} 
that $v_{\rm wf} = v_{\rm ph}(\infty)$, the {\it high-frequency\/} limit of the
phase velocity. In other words, causality is safe even if the 
low-frequency\footnote{The term ``low frequency'' in this context requires 
some clarification. We work throughout in the WKB short wavelength 
approximation  $\omega\gg R^{1/2}$  and in the limit of weak curvature
$R\ll m^2$ where $R$ is a characteristic curvature of the background 
(which can also include derivatives of the curvature) and $m$ is the electron 
mass. The frequency enters in the dimensionless ration $\omega^2R/m^4$ 
and when we talk about ``low'' and ``high'' frequency we really mean small and 
large values of this dimensionless parameter.} phase velocity $v_{\rm ph}(0)$
is superluminal provided the high-frequency limit does not exceed $c$.

This appears to remove the potential paradox associated with a superluminal
$v_{\rm ph}(0)$. However, a crucial constraint is imposed by the 
Kramers-Kronig dispersion relation\footnote{Note that we are using ``dispersion
relation'' in two different senses here.  For clarity, we will always refer
to eq.\eqref{kk} explicitly as the Kramers-Kronig or KK dispersion relation
to distinguish it from the use of the term dispersion relation to describe the
frequency dependence of the photon light-cone.} (see, 
{\it e.g.\/}~ref.~\cite{Weinberg}, chpt.~10.8)
for the refractive index, {\it viz\/}.
\EQ{
\text{Re}\,n(\infty)-\text{Re}\,n(0)=-\frac2\pi\int_0^\infty
\frac{d\omega}\omega\,\text{Im}\,n(\omega)\ .
\label{kk}
}
where $\text{Re}\,n(\omega)=1/v_\text{ph}(\omega)$.
The positivity of $\text{Im}\,n(\omega)$, which is true for an absorptive
medium and is more generally a consequence of unitarity in QFT, then 
implies that $\text{Re}\,n(\infty) < \text{Re}\,n(0)$,
i.e. $v_{\rm ph}(\infty) > v_{\rm ph}(0)$.
So, given the validity of the KK dispersion relation, a superluminal
$v_{\rm ph}(0)$ would imply a superluminal wavefront velocity 
$v_{\rm wf} = v_{\rm ph}(\infty)$ with the consequent violation of causality.

We are therefore left with three main options \cite{Shore:2007um},
each of which would have dramatic consequences for our established
ideas about quantum field theory:
\begin{quote}
{\bf Option (1)}~ 
The wavefront speed of light $v_{\rm wf} > 1$ and the
physical light cones lie outside the geometric null cones of the curved
spacetime, in apparent violation of causality.
\end{quote}
\noindent It should be noted, however, that while this would certainly 
violate causality for theories in Minkowski spacetime, it could still be 
possible for causality to be preserved in curved spacetime if the 
effective metric characterizing the physical light cones defined by 
$v_{\rm wf}$ nevertheless allow the existence of a global timelike Killing
vector field. This possible loophole exploits the general relativity notion
of ``stable causality'' \cite{HawkingEllis,Liberati:2001sd} 
and is discussed further in ref.~\cite{Shore:2003jx}.
\begin{quote}
{\bf Option (2)}~ 
Curved spacetime may behave as an optical medium exhibiting {\it
  gain\/}, {\it i.e.\/}~$\text{Im}\,n(\omega)<0$. 
\end{quote}
\noindent This possibility was explored in the context of $\L$-systems in 
atomic physics in ref.\cite{Shore:2007um}, where laser-atom interactions can
induce gain, giving rise to a negative $\text{Im}\,n(\omega)$ and superluminal 
low-frequency phase velocities while preserving $v_{\rm wf} =1$ and the
KK dispersion relation. However, the problem in extending this idea to QFT
is that the optical theorem, itself a consequence of unitarity, identifies the 
imaginary part of forward scattering amplitudes with the total cross section.
Here, $\text{Im}\,n(\omega)$ should be proportional to the cross section for
$e^+ e^-$ pair creation and therefore positive. A negative 
$\text{Im}\,n(\omega)$ would appear to violate unitarity.
\begin{quote}
{\bf Option (3)}~
The Kramers-Kronig dispersion relation \eqref{kk} is itself violated.  
Note, however, that this relation only relies on the analyticity of 
$n(\omega)$ in the upper-half plane, which is usually considered to be
a direct consequence of an
apparently fundamental axiom of local quantum field theory, 
{\it viz\/}.~{\it micro-causality\/}.
\end{quote}
\noindent Micro-causality in QFT is the requirement that the expectation
value of the commutator of field operators $\langle0|[A(x),A(y)]|0\rangle$ 
vanishes when $x$ and $y$ are spacelike separated. While this appears to be
a clear statement of what we would understand by causality at the 
quantum level, in fact its primary r\^ole in conventional QFT is as a
necessary condition for Lorentz invariance of the $S$-matrix 
(see {\it e.g.\/}~ref.~\cite{Weinberg}, chpts.~5.1,~3.5).
Since QFT in curved spacetime is only locally, and not globally, Lorentz
invariant, it is just possible there is a loophole here allowing violation
of micro-causality in curved spacetime QFT.

Despite these various caveats, unitarity, micro-causality, the identification
of light cones with geometric null cones and causality itself are all such
fundamental elements of local relativistic QFT that any one of these options
would represent a major surprise and pose a severe challenge to  
established wisdom. Nonetheless, it appears that at least one has to be true.

To understand how QED in curved spacetime is reconciled with causality, 
it is therefore necessary to perform an explicit calculation to determine
the full frequency dependence of the refractive index $n(\omega)$ in
curved spacetime. This is the technical problem which we solve in this paper.
The remarkable result is that QED chooses option (3),  
{\it viz\/}.~{\it analyticity 
is violated in curved spacetime\/}. We find that in 
the high-frequency limit, the
phase velocity always approaches $c$, so we determine $v_{\rm wf} = 1$.
Moreover, we are able to confirm that where the background gravitational
field induces pair creation, $\c \rightarrow e^+ e^-$,
$\text{Im}\,n(\omega)$ is indeed positive as required by unitarity.
However, the refractive index $n(\omega)$ is {\it not\/} analytic in the upper
half-plane, and the KK dispersion relation is modified accordingly.
One might think that this implies
a violation of microcausality, however, there is a caveat in this line
of argument which requires a more ambitious
off-shell calculation to settle definitively \cite{us}.

In order to establish this result, we have had to apply radically new
techniques to the analysis of the vacuum polarization for QED in
curved spacetime. The original Drummond-Hathrell analysis was based on
the low-energy, 
${\cal O}(R/m^2)$ effective action for QED in a curved background,
\EQ{
{\cal L}=-\frac14F_{\mu\nu}F^{\mu\nu}+\frac{\alpha}{m^2}\Big(
a R F_{\mu\nu}F^{\mu\nu} + b R_{\mu\nu}
F^{\mu\lambda}F^\nu{}_\lambda+ c R_{\mu\nu\lambda\rho}F^{\mu\nu}
F^{\lambda\rho}\Big)+\cdots\ .
}
derived using conventional heat-kernel or proper-time techniques (see,
for example, 
\cite{Gilkey:1975iq,BGVZone,BGVZtwo,Avramidi:1997jy,Barvinsky:2003rx}.
A geometric optics, or eikonal, analysis applied to this action
determines the low-frequency limit of the phase velocity. Depending on the 
spacetime, the photon trajectory and its polarization, $v_{\rm ph}(0)$ may be
superluminal \cite{Drummond:1979pp,Daniels:1993yi,Shore:2000bs}.
In subsequent work, the expansion of the effective action to all orders
in derivatives, but still at ${\cal O}(R/m^2)$, was evaluated and applied to 
the photon dispersion relation 
\cite{BGVZone,BGVZtwo,Shore:2002gw,Shore:2002gn}. 
However, as emphasized already in 
refs.~\cite{Shore:2003jx,Shore:2003zc,Shore:2002gn}, 
the derivative expansion is inadequate to find the high-frequency behaviour
of the phase velocity. The reason is that the frequency $\w$ appears in the 
{\it on-shell\/} vacuum polarization tensor only in the dimensionless ratio
${\w^2 R/ m^4}$. The high-frequency limit depends non-perturbatively 
on this parameter\footnote{Notice that here we also include derivatives of
the curvature in the generic symbol ``$R$''. In fact, in ref.~\cite{Shore:2002gn}, 
the ${\cal O}(R/m^2)$ contribution to the on-shell vacuum polarization was 
determined in the form 
$\Pi(\w) \sim \tfrac1{m^2} f({\w \over m^2}\ell\cdot D)R$, 
where $\ell\cdot D R$ represents the variation
of the curvature along the geodesic with tangent vector $\ell^\m$ and
the function $f$ is a form-factor determined from the effective
action. This behaviour, where the vacuum polarization  
depends on the curvature through its variation $\partial_u R$, where $u$ is 
a light-cone coordinate adapted to the photon's original null geodesic,
is reflected in the form of the Penrose limit for general curved spacetimes:
see Section 7.} and so is not accessible to an expansion truncated
at first order in $R/m^2$.

In this paper, we instead use the world-line formalism which can be
traced back to Feynman and Schwinger
\cite{Feynman:1950ir,Schwinger:1951xk}, 
and which has been extensively developed in recent years into a powerful tool 
for computing Green functions in QFT via path integrals for an appropriate 
1-dim world-line sigma model. 
(For a review, see {\it e.g.\/}~ref.~\cite{Schubert:2001he}.)
The power of this technique in the present context
is that it enables us to calculate the QED vacuum polarization
non-perturbatively in the frequency parameter ${\w^2 R/ m^4}$ using
saddle-point techniques. Moreover, the world-line sigma model provides
an extremely geometric interpretation of the calculation of the quantum
corrections to the vacuum polarization. In particular, we are able to give
a very direct interpretation of the origin of the Kramers-Kronig violating
poles in $n(\w)$ in terms of the general relativistic theory of null 
congruences and the relation of geodesic focusing to the Weyl and Ricci
curvatures via the Raychoudhuri equations.  

A further key insight is that to leading order in $R/m^2$, but still
exact in ${\w^2 R/ m^4}$, the relevant tidal effects of the curvature on 
photon propagation are encoded in the Penrose plane-wave limit 
\cite{Penrose,Blau2} of the spacetime expanded about the
original null geodesic traced by the photon. This is a huge
simplification, since it reduces the problem of studying photon
propagation in an arbitrary background to the much more tractable case 
of a plane wave. In fact, the Penrose limit is ideally suited to this
physical problem. As shown in ref.~\cite{Blau:2006ar}, 
where the relation with null
Fermi normal coordinates is explained, it can be extended into a systematic 
expansion in a scaling parameter which for our problem is identified as 
$R/m^2$. The Penrose expansion therefore provides us with a systematic way 
to go beyond leading order in curvature. 

The paper is organized as follows. In Section 2, we introduce the world-line 
formalism and set up the geometric sigma model and eikonal approximation.
The relation of the Penrose limit to the $R/m^2$ expansion is then explained
in detail, complemented by a power-counting analysis in the appendix.
The geometry of null congruences is introduced in Section 3, together with
the simplified symmetric plane wave background in which we perform our
detailed calculation of the refractive index. This calculation, which is the 
heart of the paper, is presented in Section 4. The interpretation of the
result for the refractive index is given in Section 5, where we plot the 
frequency dependence of $n(\w)$ and prove that asymptotically 
$v_{\rm ph}(\w) \rightarrow 1$. We also explain exactly how the existence
of conjugate points in a null congruence leads to zero modes in the sigma
model partition function, which in turn produces the KK-violating poles 
in $n(\w)$ in the upper half-plane. The implications for micro-causality 
are described in Section 6. Finally, in Section 7 we make some further 
remarks on the generality of our results for arbitrary background spacetimes 
before summarizing our conclusions in Section 8.

\section{The World-Line Formalism}

\begin{figure}[ht] 
\centerline{\includegraphics[width=2.5in]{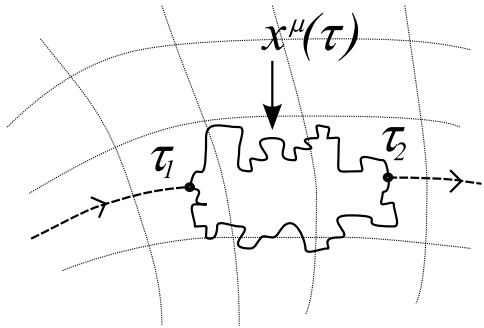}}
\caption{\footnotesize The loop $x^\mu(\tau)$ with insertions of photon vertex
  operators at $\tau_1$ and $\tau_2$.}\label{pic7}
\end{figure}
In the world-line formalism for scalar QED\footnote{Since all the conceptual 
issues we address are the same for scalars and spinors, for simplicity we 
perform explicit calculations for scalar QED in this paper. The generalization 
of the world-line formalism to spinor QED is straightforward and involves the 
addition of a further, Grassmann, field in the path integral. For ease of 
language, we still use the terms electron and positron to describe the 
scalar particles.} the 1-loop vacuum polarization is given by 
\EQ{
\Pi^\text{1-loop}=
\frac\alpha{4\pi}
\int_0^\infty\frac{dT}{T^3}\int_0^T d\tau_1\,d\tau_2\,{\cal Z}\,
\big\langle V^*_{\omega,\varepsilon_1}[x(\tau_1)]
V_{\omega,\varepsilon_2}[x(\tau_2)]\big\rangle\ .
\label{jii}
}
The loop with the photon insertions is illustrated in Figure~(\ref{pic7}).
The expectation value is calculated in the 
one-dimensional world-line sigma model involving periodic fields
$x^\mu(\tau)=x^\mu(\tau+T)$ with an action
\EQ{
S=\int_0^T d\tau\,\Big(\frac14g_{\mu\nu}(x)\dot x^\mu\dot x^\nu-m^2\Big)\ ,
}
where $m$ is the mass of the (scalar) electron and we work in Minkowski 
signature in both spacetime and on the
world-line.\footnote{This will require some appropriate $i\epsilon$
prescription. In particular, the $T$ integration contour should 
lie just below the real axis to ensure that the integral converges at 
infinity.} 
The factor ${\cal Z}$ is the partition function of
the world-line sigma model relative to flat space.\footnote{In general, 
one has to introduce ghost fields to take
account of the non-trivial measure for the fields,
$$
\int [dx^\mu(\tau)\sqrt{-\det\,g(x^\mu(\tau))}]\ ,
$$
in curved spacetime
\cite{Bastianelli:2002fv,Bastianelli:2003bg,Bastianelli:2005rc,
Kleinert:2002zm,Kleinert:2002zn}.
However, in our calculation where we work to leading
order in $R/m^2$ in a special set of coordinates 
the determinant factor is 1 to leading order.  } It is an important
detail of our calculation that ${\cal Z}$ will depend implicitly 
on $\omega$ and the insertion points $\tau_1$ and $\tau_2$.

The vertex operators have the form
\EQ{
V_{\omega,\varepsilon}[x]=\dot x^\mu A_\mu(x)\ ,
}
where $A_\mu(x)$ is the gauge connection of a photon propagating with
momentum $k$ and polarization vector $\varepsilon$.
At the one-loop level, we can impose the tree-level on-shell conditions 
for the gauge field. This means $D_\mu F^{\mu\nu}=0$ along
with the gauge condition $D_\mu A^\mu=0$. In curved spacetime, the photon
gauge field is not exactly that of a plane wave due to the effects of
curvature and in general it would be impossible to solve for the
on-shell vertex operator. However, we will work in the WKB, or short 
wavelength, approximation which is valid when
$\omega\gg R^{1/2}$.\footnote{It is important to understand that this
  notion of high frequency still allows one to expand the effective
  action in powers of $\omega$ because this latter is
  actually a function of the dimensionless ratio $\omega^2 R/m^4$
  which can be small.} 
This is the limit of geometric optics where
$A_\mu(x)$ is approximated by a
rapidly varying exponential times a much more slowly varying
polarization. Systematically, we have
\EQ{
A_\mu(x)=\big(\varepsilon_\mu(x)+\omega^{-1}B_\mu(x)
+\cdots\big)e^{i\omega\Theta(x)}\ .
\label{wkb}
}
We will need the expressions for the leading order pieces $\Theta$ and
$\varepsilon$. This will necessitate solving the on-shell conditions
to the first two non-trivial orders in the expansion in $R^{1/2}/\omega$. 
To leading order, the wave-vector $k_\m = \w \ell_\m$, where
$\ell_\mu=\partial_\mu\Theta$ is a null vector (or more
properly a null 1-form) satisfying the {\it eikonal equation\/},
\EQ{
\ell\cdot\ell\equiv g^{\mu\nu}\partial_\mu\Theta\partial_\nu\Theta=0\ .
\label{eik}
}
A solution of the eikonal equation determines a family or 
{\it congruence\/} of null
geodesics in the following way.\footnote{The congruence is not, in
  general, unique due to existence of integration constants. Later we
  will find that our results are independent of these integration constants.}
The contravariant vector field 
\EQ{
\ell^\mu(x)=\partial^\mu\Theta(x)\ ,
\label{pyu}
}
is the tangent vector to the null geodesic in the congruence
passing through the point $x^\mu$. In the particle interpretation,
$k^\mu=\omega\ell^\mu$ 
is the momentum of a photon travelling along the geodesic through
that particular point. It will turn out that the 
behaviour of the congruence will have a crucial r\^ole to play in the
resulting behaviour of the refractive index. The general relativistic
theory of null congruences is considered in detail in Section 3. 

Now we turn to the polarization vector. 
To leading order in the WKB approximation, this 
is simply orthogonal to $\ell$, {\it i.e.\/} $\varepsilon\cdot
\ell=0$. Notice that this does not determine the overall
normalization of $\varepsilon$, the {\it scalar amplitude\/}, 
which will be a space-dependent function
in general. It is useful to split $\varepsilon^\mu={\cal A}
\hat\varepsilon^\mu$, where $\hat\varepsilon^\mu$ is unit
normalized. At the next order, the WKB approximation requires that
$\hat\varepsilon^\mu$ is parallel transported along the geodesics:
\EQ{
\ell\cdot D\,\hat\varepsilon^\mu=0\ .
\label{uiu}
}
The remaining part, the scalar amplitude ${\cal A}$, satisfies 
\EQ{
\ell\cdot D\,\log{\cal A}=-\frac12D\cdot\ell\ .
\label{did}
}
Eqs.~\eqref{uiu} and \eqref{did} are equivalent to
\EQ{
\ell\cdot D\,\varepsilon^\mu=-\frac12\varepsilon^\mu D\cdot\ell\ .
\label{pip}
}
Since the polarization vector is defined up to an additive amount of
$k$, there are two linearly independent polarizations
$\varepsilon_i(x)$, $i=1,2$. 

Since there are two polarization states, the one-loop vacuum
polarization is actually a $2\times2$ matrix
\SP{
\Pi^\text{1-loop}_{ij}
&=\frac\alpha{4\pi}\int_0^\infty\frac{dT}{T^3}
\int_0^Td\tau_1\,d\tau_2\,{\cal Z}\\
&\times 
\Big\langle \varepsilon_i[x(\tau_1)]\cdot\dot x(\tau_1)
e^{-i\omega\Theta[x(\tau_1)]}\,
\varepsilon_j[x(\tau_2)]\cdot
\dot x(\tau_2)e^{i\omega\Theta[x(\tau_2)]}
\Big\rangle\ .
\label{poo}
}
In order for this to be properly defined we must specify how to
deal with the zero mode of $x^\mu(\tau)$ in the world-line sigma
model. Two distinct -- but ultimately equivalent -- methods for dealing
with the zero mode have been proposed in the literature
\cite{Bastianelli:2002fv,Bastianelli:2003bg,Bastianelli:2005rc,
Kleinert:2002zm,Kleinert:2002zn}.
In the first,
the position of one particular point on the loop is defined as the zero mode,
while in the other, the ``string inspired'' definition, the zero mode
is defined as the average position  of the loop:
\EQ{
x_0^\mu=\frac1T\int_0^T d\tau\,x^\mu(\tau)\ .
\label{con}
}
We will use this latter definition since it leads to a much simpler
formalism. Since we are effectively calculating an on-shell term in an
effective action, the integral over the zero mode is simply 
excluded from the functional integral. In other words, our
world-line sigma model does not include an integral over $x_0^\mu$
which one should think of as being a fixed point in spacetime. Since
in curved spacetime there is in general no translational symmetry, the
one-loop correction $\Pi^\text{1-loop}_{ij}(x_0)$ 
will depend explicitly on $x_0^\mu$. We will always choose
coordinates for which $x_0^\mu=0$, in which case we implicitly impose
the constraint
\EQ{
\int_0^T d\tau\,x^\mu(\tau)=0
\label{con2}
}
on the sigma model fields. 
The advantage of using the string inspired method is that there is 
translational symmetry on the world-line loop. This allows us to
fix $\tau_1=0$. We will then take
$\tau_1=\xi T$, $0\leq\xi\leq 1$, 
and replace the two integrals over $\tau_1$ and $\tau_2$ by a single integral
over the variable $\xi$.

A key ingredient in our analysis is that in the
limit of weak curvature $R\ll m^2$, the sigma model based on the general
metric $g_{\mu\nu}$ can be approximated by the metric in a cylindrical
neighbourhood of the geodesic in the null congruence 
that passes through $x_0^\mu=0$. We
will call this particular geodesic $\gamma$. The metric in the
neighbourhood of $\gamma$ arises in a very particular way 
known as the Penrose limit \cite{Penrose}. Exactly how this limit
arises is rather remarkable and means that the vacuum
polarization and refractive index 
is only sensitive to the Penrose limit of the original metric.
It should be noted that the Penrose limit
captures the global behaviour of the original metric all the way along the
geodesic $\gamma$.

\begin{figure}[ht] 
\centerline{\includegraphics[width=3.5in]{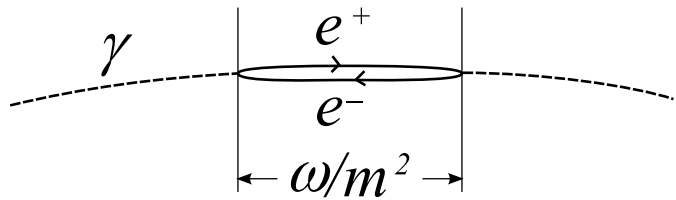}}
\caption{\footnotesize The classical saddle point solution consists
of a squashed loop which follows the geodesic $\gamma$.
The length of the loop is $\sim\omega/m^2$ which represents a
potentially interesting UV-IR mixing effect.}\label{pic10}
\end{figure}

Now notice that the exponential pieces of the vertex
operators in \eqref{jii} act as source terms and so the complete action
including these is
\EQ{
S=-T+\frac{m^2}{4T}\int_0^1d\tau\,g_{\mu\nu}(x)\dot x^\mu\dot x^\nu-
\omega\Theta[x(\xi)]+\omega\Theta[x(0)]\ .
\label{iuu}
}
Here, we have scaled  $\tau\to T\tau$ and then $T\to T/m^2$, so that 
$\tau$ runs from 0 to 1. $T$ is now dimensionless and $1/m^2$
plays the r\^ole of a conventional coupling constant. 
In fact, the effective coupling constant
is actually the dimensionless ratio $R/m^2$,
where $R$ is a typical curvature scale. So when 
$R/m^2$ is small we can perform a perturbative expansion in 
the world-line sigma model. As is usual in a perturbative analysis,
it is useful 
to re-scale the ``fields'' $x^\mu(\tau)$ appropriately 
in order to remove the overall factor of $m^2/T$. The coupling then
re-appears in vertices. However, this re-scaling must be
done in a clever way. The reason is that the classical saddle-point
solution following from \eqref{iuu} is not simply the constant configuration
$x^\mu(\tau)=x_0^\mu=0$ because the sources inject world-line momentum
into, and out of, the system. It is not difficult to guess what the classical
saddle-point solution will be because the classical equation of motion
that follows from \eqref{iuu} is just the geodesic equation for
$x^\mu(\tau)$ with delta-function sources at $\tau=0$ and $\xi$. The
solution consists of an electron and positron pair produced at
$\tau=0\equiv 1$ which propagate along the photon geodesic $\gamma$, 
with the electron going from $\tau=0$ to $\tau=\xi$ and the positron 
from $\tau=1$ to $\tau=\xi$, before
annihilating at $\tau=\xi$ back into a photon which then continues
along $\gamma$. In other words, the classical loop is squashed onto
the geodesic $\gamma$ as illustrated in Figure~(\ref{pic10}).
We will find the explicit solution for this classical loop shortly.

As we have said, 
the fact that there is a non-trivial classical solution around which
the perturbative expansion is performed means that the re-scaling of the 
fields must be done in an appropriate way. The problem is solved by 
choosing from the outset a set of coordinates which are adapted to the null 
congruence containing $\gamma$. These coordinates $(u,\Theta,Y^a)$,
$a=1,2$, are known as {\it Rosen coordinates\/}. They  
include two null coordinates: $u$, the affine parameter
along the geodesics and $\Theta$, the solution of the eikonal
equation \eqref{eik}. As explained in ref.~\cite{Blau2}, the full metric
$g_{\mu\nu}$ around $\gamma$ can always be brought into the form
\EQ{
ds^2=2du\, d\Theta-C(u,\Theta,Y^a)d\Theta^2-2C_a(u,\Theta,Y^b)
dY^a\,d\Theta -C_{ab}(u,\Theta,Y^c)dY^a\,dY^b\ .
\label{mrosen}
}
It is manifest that $d\Theta$ is a null 1-form. The null congruence
has a simple description as the curves  $(u,\Theta_0,Y^a_0)$ for fixed 
values of the transverse coordinates $(\Theta_0,Y^a_0)$. The geodesic
$\gamma$ is the particular member  $(u,0,0,0)$.
It should not be surprising that the
Rosen coordinates are singular at the {\it caustics\/} of the
congruence. These are points where members of the congruence
intersect and will be described in detail in the next section.

With the form \eqref{mrosen} of the metric, one finds that the classical
equations of motion of the sigma model action \eqref{iuu} have a
solution with $Y^a=\Theta=0$ where $u(\tau)$ satisfies 
\EQ{
\ddot u=-\frac{2\omega T}{m^2}\delta(\tau-\xi)+
\frac{2\omega T}{m^2}\delta(\tau)\ .
\label{hyq}
}
More general solutions with constant but non-vanishing $(\Theta,Y^a)$ 
are ruled out by the constraint \eqref{con2}. The solution
of \eqref{hyq} is
\EQ{
\tilde u(\tau)=-u_0+\begin{cases}2\omega T(1-\xi)\tau/m^2 & 0\leq \tau\leq
  \xi\\ 2\omega T\xi(1-\tau)/m^2 & \xi\leq\tau\leq 1\ .\end{cases}
\label{gsol}
}
where the constant 
\EQ{
u_0=\omega T\xi(1-\xi)/m^2
} 
ensures that the constraint
\eqref{con2} is satisfied. The solution describes a loop which is 
squashed down onto the geodesic $\gamma$ as illustrated in 
Figure~(\ref{pic10}). The electron and positron have to
move with different world-line velocities 
in order to accommodate the fact that in general $\xi$ is not equal to
$\tfrac12$. In Section 5, 
we explain how for particular values of $T$ there are more
general classical saddle-point solutions which are consistent with
\eqref{con2}. However, the solution we
have described is the only one that exists for generic values of $T$.

What is intriguing about this picture is that the classical loop, 
which has an affine parameter length proportional 
to $L\sim\omega/m^2$, actually gets bigger as the frequency is increased. 
The reason is that higher frequency leads to bigger impulses and hence 
longer loops. This is an interesting 
example of the kind of UV-IR mixing that is seen in other
contexts, such as non-commutative field theories or high
energy string scattering. Whether the occurrence here is hinting at 
something deeper deserves to be investigated in more detail. However,
what it will mean is that the higher frequencies will probe global
aspects of the spacetime rather than shorter distance scales as
our intuition might have suggested.

Now that we have defined the Rosen coordinates and found the classical
saddle-point solution, we are in a position to set up the
perturbative expansion. The idea is to scale the transverse
coordinates $\Theta$ and $Y^i$ in order to remove the factor of $m^2/T$
in front of the action. The affine coordinate $u$, on the other
hand, will be left alone since the classical solution
$\tilde u(\tau)$ is by definition 
of zeroth order in perturbation theory. The appropriate scalings are
precisely those needed to define the Penrose limit \cite{Penrose} -- in 
particular we closely follow the discussion in \cite{Blau2}.
The Penrose limit involves first a boost 
\EQ{
(u,\Theta,Y^a)\longrightarrow(\lambda^{-1}u,\lambda\Theta,Y^a)\ ,
}
where $\lambda=T^{1/2}/m$, 
and then a uniform re-scaling of the coordinates
\EQ{
(u,\Theta,Y^a)\longrightarrow(\lambda u,\lambda\Theta,
\lambda Y^a)\ .
}
As argued above, 
it is important that the null coordinate along the geodesic $u$ is
not affected by the combination of the boost and re-scaling; indeed, overall
\EQ{
(u,\Theta,Y^a)\longrightarrow(u,\lambda^2\Theta,\lambda Y^a)\ .
\label{ps}
}
After these re-scalings, 
the sigma model action \eqref{iuu} becomes
\SP{
S=&-T+
\frac14\int_0^1d\tau\,\Big[2\dot u\,\dot\Theta-\lambda^2
C(u,\lambda^2\Theta,\lambda Y^a)\dot\Theta^2 
\\ &-2\lambda C_a(u,\lambda^2\Theta,\lambda Y^b)
\dot Y^a\,\dot\Theta-C_{ab}(u,\lambda^2\Theta,\lambda 
Y^c)\dot Y^a\,\dot 
Y^b\Big]\\ &-\frac{\omega T}{m^2}
\Theta(\xi)+\frac{\omega T}{m^2}\Theta(0)\ .
} 
In the limit $R\ll m^2$, we expand in powers of $\lambda=T^{1/2}/m$
and ignore terms of  ${\cal O}(\lambda)$:
\EQ{
S=-T+\frac14\int_0^1d\tau\,\Big[2\dot u\,\dot\Theta
-C_{ab}(u,0,0)\dot Y^a\,\dot Y^b\Big]-\frac{\omega
  T}{m^2}\Theta(\xi)+
\frac{\omega T}{m^2}\Theta(0)+\cdots\ .
}
The leading order piece is precisely the Penrose limit of the original
metric in Rosen coordinates. Notice that we must keep the source
terms because the combination $\omega T/m^2$, or more precisely the
dimensionless ratio $\omega R^{1/2}/m^2$, can be large. However, there
is a further simplifying feature: 
once we have shifted the ``field'' about the classical solution
$u(\tau)\to\tilde u(\tau)+u(\tau)$, it is clear that there are
no Feynman graphs without external $\Theta$ lines that involve the
vertices 
$\partial_u^n C_{ab}(\tilde u,0,0)u^n\,\dot Y^a\,\dot Y^b$, 
$n\geq1$; hence, we can simply
replace $C_{ab}(\tilde u+u,0,0)$ consistently with the background
expression  $C_{ab}(\tilde u,0,0)$. This means that the resulting
sigma model is Gaussian to leading order in $R/m^2$:
\EQ{
S^{(2)}=\frac14\int_0^1d\tau\,\Big[2\dot u\,\dot\Theta
-C_{ab}(\tilde u,0,0)\dot Y^a\,\dot Y^b\Big]
\longrightarrow -\frac14\int_0^1d\tau\,
C_{ab}(\tilde u,0,0)\dot Y^a\,\dot Y^b
\ ,
\label{gas}
}
where finally we have dropped the $\dot u\,\dot\Theta$
piece since it is just the same as in flat 
space and the functional integral is normalized
relative to flat space. This means that 
all the non-trivial curvature dependence
lies in the $Y^a$ subspace transverse to the geodesic.\footnote{An
alternative proof of this result which relies only on conventional
power counting arguments and does not rely on any {\it a
  priori\/}~knowledge of the Penrose limit is provided in Appendix A.}

It turns out that the Rosen coordinates are actually not the most
convenient coordinates with which 
to perform explicit calculations. For this,
we prefer {\it Brinkmann coordinates\/} $(u,v,y^i)$.
To define these, we first introduce a  
 ``zweibein'' in the subspace of the $Y^a$:
\EQ{
C_{ab}(u)=\delta_{ij}E^i{}_a(u)E^j{}_b(u)\ ,
}  
with inverse $E^a{}_i$. This quantity is subject to the 
condition that 
\EQ{
\Omega_{ij}\equiv\frac{dE_{ia}}{du}E^a{}_j
}
is a symmetric matrix.\footnote{Notice that $i$ and $j$ are raised and
  lowered in this Euclidean 
2d subspace by $\delta_{ij}$ and not $-\delta_{ij}$.}
Then the affine coordinate $u$ is common to
both systems, while
\EQ{
y^i=E^i{}_aY^a\ ,\qquad
v=\Theta+\frac12\frac{dE_{ia}}{du}E^i{}_bY^aY^b\ .
}
Notice that the Brinkmann coordinates are homogeneous under the 
scaling \eqref{ps}:
\EQ{
(u,v,y^i)\longrightarrow(u,\lambda^2v,\lambda y^i)\ .
\label{ps2}
}
In Brinkmann coordinates, the metric takes the form
\EQ{
ds^2=2du\,dv+h_{ij}(u)y^i\,y^j\,du^2-dy^{i2}\ ,
\label{spwmetric}
}
where the quadratic form is
\EQ{
h_{ij}(u)=-\frac{d^2E_{ia}}{du^2}E^a{}_j\ .
}
We have introduced these coordinates at the level of the Penrose
limit. However, they have a more general definition for an arbitrary
metric and geodesic. They are in fact {\it
  Fermi normal coordinates\/}. These are ``normal'' in the same sense 
as the more common Riemann normal coordinates, but
in this case they are associated to the geodesic curve $\gamma$
rather than to a single point. 
This description of Brinkmann coordinates as
Fermi normal coordinates and their relation to Rosen coordinates and
the Penrose limit is
described in detail in ref.\cite{Blau:2006ar}. In particular, this
reference gives the
$\lambda$ expansion of the metric in null Fermi normal coordinates to
${\cal O}(\lambda^2)$. To ${\cal O}(\lambda)$ this is
\SP{
ds^2&=2du\,dv-R_{iuju}y^i\,y^j\,du^2-dy^{i2}\\ &+
\lambda\Big[-2R_{uiuv}y^iv\,du^2-\tfrac43R_{uijk}y^iy^j\,du\,dy^k-
\tfrac13R_{uiuj;k}y^iy^jy^k\,du^2\Big]+{\cal O}(\lambda^2)\ ,
\label{penexp}
}
which is consistent with \eqref{spwmetric} since
$R_{iuju}=-h_{ij}$ for a plane wave. 
It is worth pointing out
that Brinkmann coordinates, unlike Rosen coordinates, are not singular
at the caustics of the null congruence. One can say that Fermi normal
coordinates (Brinkmann coordinates) are naturally associated to a
single geodesic $\gamma$ whereas Rosen coordinates are naturally
associated to a congruence containing $\gamma$.

In Brinkmann coordinates, the Gaussian action \eqref{gas} for the
transverse coordinates becomes
\EQ{
S^{(2)}=-\frac14\int_0^1d\tau\,\big(\dot y^{i2}-
\dot{\tilde u}^2h_{ij}(\tilde u)y^i\,y^j\big)
+\frac{\omega T}{2m^2}\Omega_{ij}y^iy^j\Big|_{\tau=\xi}-
\frac{\omega
T}{2m^2}\Omega_{ij}y^iy^j\Big|_{\tau=0}\ ,
\label{gas2}
}
where the world-line velocity along the loop is
\EQ{
\dot{\tilde u}
(\tau)=\begin{cases}2\omega T(1-\xi)/m^2 & 0\leq \tau\leq
  \xi\\ -2\omega T\xi/m^2 & \xi\leq\tau\leq 1\ .\end{cases}
}
Although \eqref{gas2} looks more complicated than \eqref{gas}, it is
actually more useful for explicit calculations.

\section{The Symmetric Plane Wave and Null Congruences}

The analysis above shows that photon propagation in a completely general
curved spacetime is governed to one-loop order by the Penrose limit for
the metric in a neighbourhood of the original null geodesic. The complete
one-loop vacuum polarization and photon dispersion relation can therefore
be determined without loss-of-generality by working in a plane 
wave background.

In Section 7, we briefly discuss the Penrose limits of spacetimes
of special physical interest, such as de Sitter and Schwarzschild, and see
how known results for low-frequency photon propagation in these spacetimes are 
recovered as special properties of the Penrose limit. For the rest of this 
paper, however, we specialize to the simplest example of a plane wave -- the
symmetric plane wave \cite{Blau2}.  In this background, we can evaluate the
non-perturbative 
frequency dependence of the vacuum polarization explicitly. In doing 
so, we discover many surprising features of the dispersion relation that will 
hold in general.

The symmetric plane wave metric is given in Brinkmann coordinates by
\eqref{spwmetric}, with the restriction that $h_{ij}$ is independent of $u$.
This metric is {\it locally symmetric\/} in the sense that the Riemann tensor
is covariantly constant, $D_\l R_{\m\n\r\s} = 0$, and can be realized as
a homogeneous space $G/H$ with isometry group $G$.\footnote{Notice that,
contrary to the implication in ref.~\cite{Shore:2002gn,Shore:2007um}, 
the condition that the Riemann tensor is covariantly constant only 
implies that the spacetime is {\it locally\/} symmetric, and not necessarily maximally symmetric \cite{Avramidi:1997jy,Blau2}. A maximally symmetric space has 
$R_{\m\n\r\s} = {1\over12}R (g_{\m\r} g_{\n\s} - g_{\m\s} g_{\n\r})$ 
and does not have the required anisotropy for the
vacuum polarization to modify the speed of light.} With no loss of generality,
we can choose a basis for the transverse coordinates in which $h_{ij}$ is
diagonal:
\EQ{
h_{ij} y^i y^j = \s_1^2( y^1)^2 + \s_2^2( y^2)^2\ .
\label{ca}
}
The sign of these coefficients plays a crucial role, so we allow the 
$\s_i$ themselves to be purely real or purely imaginary. 

For a general plane-wave metric, the only non-vanishing components of the
Riemann tensor (up to symmetries) are
\EQ{
R_{u i u j} = - h_{ij}(u)\ .
\label{cb}
}
So for the symmetric plane wave, we have simply
\SP{
R_{uu} &=\s_1^2 + \s_2^2\ ,\\
R_{ui u i} &= - \s_i^2
\label{cc}
}
and for the Weyl tensor,
\EQ{
C_{u i u i} = -\s_i^2 +{1\over2} \sum_{j=1}^2 \s_j^2\ .
\label{cd}
}
The {\it null energy condition\/}, {\it viz\/}.~$T_{\m\n}k^\m k^\n \ge 0$ with 
$k^\m$ a
null vector, reduces here to $T_{uu} \ge 0$, so from Einstein's equation
we require $R_{uu} = \s_1^2 + \s_2^2 \ge 0$.  It follows that at least one 
of the $\s_i$ must be real (we will always choose this to be 
$\sigma_1$).  Special choices of the $\s_i$ allow the symmetric 
plane wave to be either Ricci flat ($\s_1 = \pm i \s_2$) or conformally
flat ($\s_1 = \pm \s_2$). The Ricci flat case is the vacuum
gravitational wave.

While, as we saw in the last section, the original null geodesic $\c$
(with $\ell = \partial_u$) defines the classical solution in the 
world-line path integral, in order to evaluate the fluctuations we also 
need the eikonal phase and wave-vector for deviations from $\c$ itself.
We therefore need to study the congruence of null geodesics in the
neighbourhood of $\c$ in Brinkmann coordinates. 
We first do this explicitly for the symmetric
plane wave background, then explain how the key features are described in 
the general theory of null congruences.

The geodesic equations for the symmetric plane wave \eqref{spwmetric},
\eqref{ca} are:
\SP{
\ddot u&= 0\ ,\\
\ddot v + 2 \dot u \sum_{i=1}^2 \s_i^2 y^i \dot y^i &= 0 \ ,\\
\ddot y^i + {\dot u}^2 \s_i^2 y^i &= 0\ . 
\label{ce}
}
We can therefore take $u$ itself to be the affine parameter and, with the
appropriate choice of boundary conditions, define the null congruence in the
neighbourhood of, and including, $\c$ as:
\SP{
v &=\Theta - {1\over2} \sum_{i=1}^2 \s_i \tan(\s_i u + a_i) y^{i2} \ ,\\
y^i &= Y^i \cos(\s_i u + a_i)\ . 
\label{cf}
}
The constants $\Theta$ and $Y^i$ are nothing other than the
Rosen coordinates for the symmetric plane wave. In fact, in Rosen
coordinates the symmetric plane wave metric is
\EQ{
ds^2=2du\,d\Theta-\sum_{i=1}^2\cos^2(\sigma_iu+a_i)dY^{i2}\ .
}
The integration constants $a_i$ can be thought of 
as redundancies in the definition of the null congruence and the
associated Rosen coordinates; in particular, they determine the position of
the caustics. Given this, we have
\SP{
E^i{}_a&=\delta_{ia}\cos(\sigma_iu+a_i)\ ,\\
E^a{}_i&=\delta_{ia}\sec(\sigma_iu+a_i)\ ,\\
\Omega_{ij}&=-\delta_{ij}\sigma_i\tan(\sigma_iu+a_i)
}
and it is immediate that the eikonal phase is 
\EQ{
\Theta(x) = v + {1\over2}\sum_{i=1}^2 \s_i \tan(\s_i u + a_i) y^{i2}\ .
\label{cg}
}
The tangent vector to the congruence, defined as 
$\ell^\m = g^{\m\n} \partial_\n \Theta$, is therefore
\EQ{
\ell=\partial_u + {1\over2} \sum_{i=1}^2\Big\{\s_i^2 
\bigl(\tan^2(\s_i u + a_i)-1\bigr)
y^{i2}\ \partial_v - \s_i \tan(\s_i u + a_i) y^i \partial_i\Big\}\ .
\label{ch}
}
The polarization vectors are orthogonal
to this tangent vector, $\ell\cdot\varepsilon_{i} = 0$, and are
further constrained by \eqref{pip}. 
Solving \eqref{uiu} for the normalized polarization (one-form)
yields\footnote{The one-form is exactly what appears in the vertex
  operator via $\varepsilon_{i\mu}\dot x^\mu$.} 
\EQ{
\hat\varepsilon_i =dy^i+\s_i \tan(\s_i u + a_i)y^idu\ .
\label{ci}
}
The scalar amplitude ${\cal A}$ is determined by the parallel transport 
equation \eqref{did}, from which we readily find (normalizing
so that ${\cal A}(0) = 1$)
\EQ{
{\cal A} =\prod_{i=1}^2  \sqrt{\cos a_i \over \cos(\s_i u + a_i)} 
\label{cj}
}

The null congruence in the symmetric plane wave background displays a number
of features which play a crucial role in the analysis of the refractive index.
They are best exhibited by considering the Raychoudhuri equation, which 
expresses
the behaviour of the congruence in terms of the {\it optical
  scalars\/}, {\it viz\/}.~the
expansion $\hat\theta$, shear $\hat\s$ and twist $\hat\omega$. These are 
defined in terms of the covariant derivative of the tangent vector 
as \cite{Chandra}:
\SP{
\hat\theta &=\tfrac12D_\m \ell^\m\ ,\\
\hat\s &=\sqrt{\tfrac12D_{(\m}\ell_{\n)} D^\m \ell^\n - 
\hat\theta^2}\ ,\\
\hat\w &= \sqrt{\tfrac12D_{[\m}\ell_{\n]} D^\m \ell^\n}\ .
\label{ck}
}
The Raychoudhuri equations describe the variation of the optical scalars along
the congruence:
\SP{
\partial_u{\hat\theta}&=-\hat\theta^2 - \hat\s^2 + \hat\w^2 - \Phi_{00}\ ,\\
\partial_u{\hat\s}&= -2 \hat\theta \hat\s  - |\Psi_0|\ .
\label{cl}
}
(We will not need the equation for the twist.)
Here, we have introduced the Newman-Penrose notation 
(see, {\it e.g.\/}~ref.~\cite{Chandra}) for the components 
of the Ricci and Weyl tensors: $\Phi_{00} = {1\over2} R_{\m\n}\ell^\m \ell^\n$,
~~$\Psi_0 = C_{\m\r\n\s}\ell^\m \ell^\n m^\r m^\s$.\footnote{For the symmetric
plane wave, the Newman-Penrose null tetrad basis $\ell^\m, n^\m, m^\m,
\bar m^\m$
comprises $\ell$ as in eq.~\eqref{ch}, $n = \partial_v$, and 
$m= {1\over\sqrt2}(\varepsilon_1 + i \varepsilon_2)$.  
The basis vectors satisfy
$\ell\cdot n = 1$, $m\cdot\bar m = -1$ and the metric can be expressed as
$g_{\m\n} = 2 \bigl( \ell_{(\m} n_{\n)} - m_{(\m} \bar m_{\n)}\bigr)$. 
Since eqs.~\eqref{cc} and \eqref{cd} are the only non-vanishing components of 
the Ricci
and Weyl tensors, it follows that $\Phi_{00} = {1\over2} R_{uu} =
\tfrac12(\s_1^2 + \s_2^2)$, ~$\Psi_0 = {1\over2}(C_{u 1 u 1} -
C_{u 2 u 2}) = {1\over2}(\s_2^2 - \s_1^2)$.}  As demonstrated in
refs.\cite{Shore:1995fz}, 
the effect of vacuum polarization on low-frequency photon
propagation is also governed by the two curvature scalars $\Phi_{00}$ and
$\Psi_0$. Indeed, many interesting results such as the polarization sum rule 
and horizon theorem \cite{Shore:1995fz,Gibbons:2000xe}
are due directly to special properties of $\Phi_{00}$ and
$\Psi_0$. As we now show, they also play a key r\^ole in the world-line 
formalism in determining the nature of the full dispersion relation.

By its definition as a gradient field, it is clear that $D_{[\m}\ell_{\n]} = 0$
so the null congruence is twist-free $\hat\omega=0$.  
The remaining Raychoudhuri equations can then be rewritten as
\SP{
\partial_u({\hat\theta} + {\hat\s}) &= - (\hat\theta + \hat\s)^2
-\Phi_{00} -|\Psi_0|\ ,\\
\partial_u({\hat\theta} - {\hat\s}) &=- (\hat\theta - \hat\s)^2
- \Phi_{00} +|\Psi_0|\ .
\label{cm}
}
The effect of expansion and shear is easily visualized by the effect on a 
circular cross-section of the null congruence as the affine parameter $u$
is varied: the expansion $\hat\theta$ gives a uniform expansion whereas the
shear $\hat\s$ produces a squashing with expansion along one transverse axis
and compression along the other. The combinations $\hat\theta \pm \hat\s$ 
therefore describe the focusing or defocusing of the null rays in the two
orthogonal transverse axes.

We can therefore divide the symmetric plane wave 
spacetimes into two classes, depending on the
signs of $\Phi_{00} \pm |\Psi_0|$.
  A Type I spacetime, where $\Phi_{00} \pm
|\Psi_0|$ are both positive, has focusing in both directions, whereas Type II,
where $\Phi_{00} \pm \Psi_0$ have opposite signs, has one focusing and
one defocusing direction. Note, however, that there is no ``Type III'' with
both directions defocusing, since the null-energy condition requires 
$\Phi_{00} \ge 0$.

\begin{figure}[ht] 
\centerline{(a)\includegraphics[width=3in]{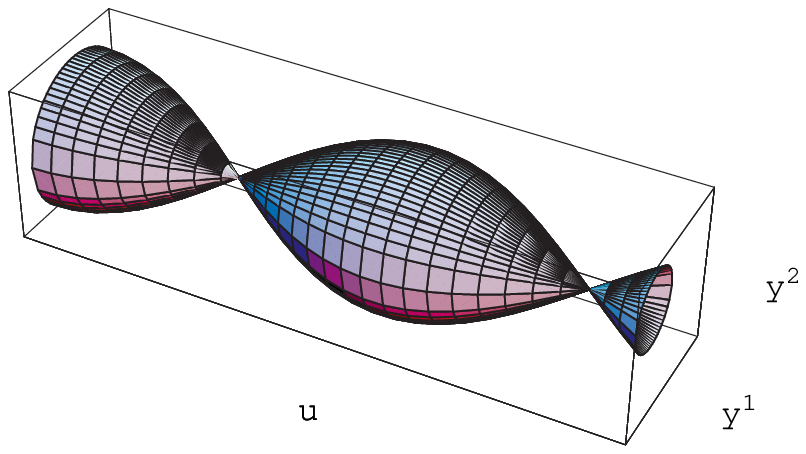}
\hspace{1cm}(b)\includegraphics[width=2in]{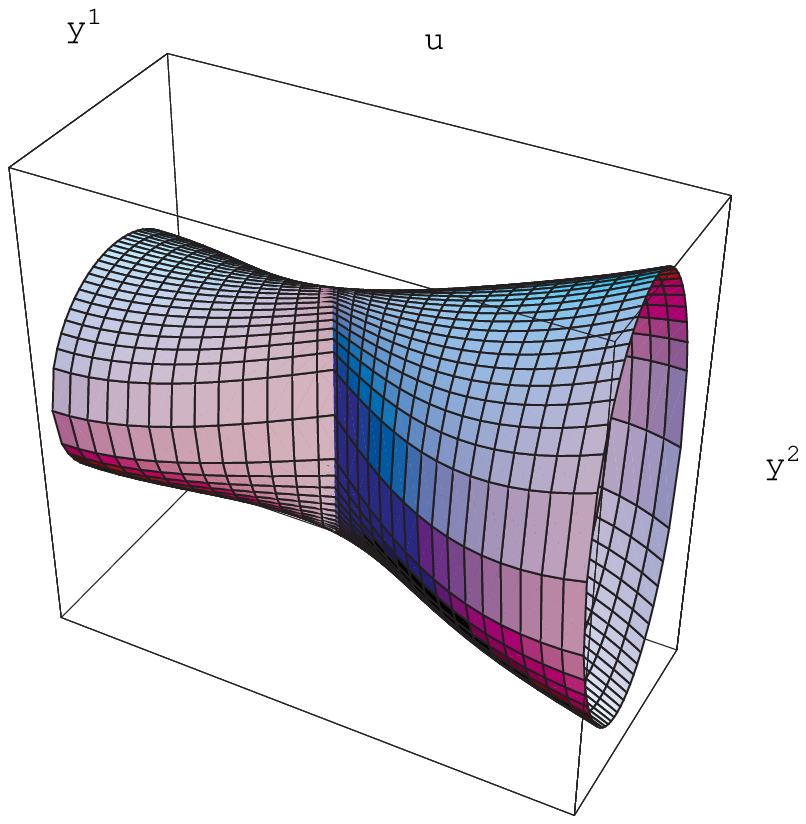}}
\caption{\footnotesize (a) Type I null congruence with the special choice
  $\sigma_1=\sigma_2$ and $a_1=a_2$ so that the caustics in both
  directions coincide as focal points. (b) Type II null congruence
  showing one focusing and one defocusing direction.} 
\label{focus}
\end{figure}

For the symmetric plane wave, the focusing or defocusing of the geodesics
is controlled by eq.\eqref{cf}, $y^i = Y^i \cos(\s_i u + a_i)$. 
Type I therefore
corresponds to $\s_1$ and $\s_2$ both real, whereas in Type II, $\s_1$ is
real and $\s_2$
is pure imaginary. The behaviour of the congruence in these two cases is
illustrated in Figure~(\ref{focus}).

To see this explicitly in terms of the Raychoudhuri equations, note first that
the curvature scalars $\Phi_{00} - |\Psi_0| = \s_1^2$, 
$\Phi_{00} + |\Psi_0| = \s_2^2$~ are simply the eigenvalues of $h_{ij}$.
The optical scalars are
\SP{
\hat\theta &= 
- {1\over2}\bigl(\s_1 \tan(\s_1 u + a_1) + \s_2 \tan(\s_2 u + a_2) \bigr)\
,\\ \hat\s &=
{1\over2}\bigl(\s_1 \tan(\s_1 u + a_1) - \s_2 \tan(\s_2 u + a_2) \bigr) 
\label{cn}
}
and we easily verify
\SP{
\partial_u \hat\theta&=\hat\theta^2 - \hat\s^2 - {1\over2}(\s_1^2 + \s_2^2)
\ ,\\
\partial_u \hat\s &= - 2 \hat\theta \hat\s + {1\over2} (\s_1^2 -
\s_2^2)\ .
\label{co}
}

 It is clear that provided the geodesics are complete, those in a focusing 
direction will eventually cross. In the symmetric plane wave example, 
with $y^i =Y^i \cos(\s_i u + a_i)$, these ``caustics'' 
occur when the affine parameter $\s_iu =\pi(n+\tfrac12)-a_i$,
$n\in{\bf Z}$. At a caustic, the amplitude
factor ${\cal A}$ in \eqref{cj} diverges and correspondingly 
the Rosen coordinates are not well defined.
In fact, the existence of 
{\it conjugate points\/}, {\it i.e.\/}~points $p$ and $q$ 
on a geodesic $\gamma$ that can
be joined by geodesics infinitesimally close to
$\gamma$, is generic in spacetimes satisfying the null energy
condition.\footnote{This does not necessarily mean that the
  conjugate points are joined by more than one actual geodesic,
  only that an infinitesimal deformation of $\gamma$ exists. Later we
  shall see that the existence of conjugate points relies on the
  existence of zero modes of a linear problem. Conversely, the
  existence of a geodesic other than $\gamma$ joining $p$ and $q$
  does not necessarily mean that $p$ and $q$ are conjugate 
\cite{HawkingEllis,Wald:1984rg}.} 
The result is summarized in the following theorem 
\cite{HawkingEllis,Wald:1984rg}:

\begin{quote}
{\bf Theorem:} if a spacetime satisfies the ``null generic condition''
({\it i.e.\/}~every
null geodesic has at least one point where either
\EQ{ R_{\m\n}\ell^\m \ell^\n \ne 0\qquad\text{  or  }
\qquad\ell_{[\l} C_{\m]\r\n[\s} \ell_{\t]}\ell^\rho \ell^\nu \ne 0,
} or equivalently
$\Phi_{00} \ne 0$ or $\Psi_0 \ne 0$) and the null energy condition, then every
complete null geodesic possesses a pair of conjugate points.
\end{quote}

The existence of conjugate points will turn out to be 
crucial in the world-line sigma model formalism. 
It means that for certain values of $T$ (for a given $\omega$), such
that $u=\pm u_0$ are conjugate points, in the Penrose limit around
the geodesic, there exists a family of classical solutions 
corresponding to the different geodesic paths between
the conjugate points.\footnote{Whether these deformed geodesics become
  actual geodesics is the question as to whether they lift from the
  Penrose limit to the full metric.}
This implies the existence of zero modes which, as explained in Section 5,
ultimately controls the location of singularities of the refractive index
in the complex $\w$ plane and is the key to understanding the violation
of the conventional Kramers-Kronig dispersion relation and the fate of
micro-causality. 
  
\section{World-line Calculation of the Refractive Index}

In this section, we calculate the vacuum polarization and refractive
index explicitly for a symmetric plane wave. As we mentioned at the
end of Section 2, the explicit calculations are best performed in
Brinkmann coordinates. We will need the expressions for $\Theta$ and
$\varepsilon_i$ for the symmetric plane wave background: these are in
eqs.\eqref{cg}, 
\eqref{ci} and \eqref{cj}. From these, we have the following explicit
expression for the 
vertex operator\footnote{Notice that at leading
order in $R/m^2$ we are at liberty to replace $u(\tau)$ by its
classical value $\tilde u(\tau)$. The argument is identical to the
one given in Section 2.}
\SP{
V_{\omega,\varepsilon_i}[x^\mu(\tau)]=&
\big(\dot
y^i+\sigma_i\tan(\sigma_i\tilde u+a_i)\dot{\tilde u}y^i\big)
\prod_{j=1}^2\sqrt{\frac{\cos a_i}{\cos(\sigma_j\tilde u+a_j)}}\\
&\times\exp i\omega\Big[v+\frac12\sum_{j=1}^2
\sigma_j\tan(\sigma_j\tilde u+a_j)y^{j2}\Big]\ .
\label{loi}
}
The Gaussian action for the transverse coordinates, 
including the source terms \eqref{gas2}, is
\SP{
S^{(2)}=&\sum_{i=1}^2\Big\{-\frac1{4}\int_0^1d\tau\,\big(
\dot y^{i2}-\dot{\tilde u}^2\sigma_i^2y^{i2}\big)\\ &
-\frac{\omega T\sigma_i}{2m^2}
\big(\tan(\sigma_iu_0+a_i)y^i(\xi)^2+\tan(\sigma_iu_0-a_i)y^i(0)^2\big)
\Big\}\ .
}
Notice that the $y^i$ fluctuations are completely decoupled. The
measure for the field $x^\mu(\tau)$ is covariant and so 
includes the factor $\sqrt{-\det\,g[x(\tau)]}$ which can be exponentiated by
introducing appropriate ghosts 
\cite{Bastianelli:2002fv,Bastianelli:2003bg,Bastianelli:2005rc,
Kleinert:2002zm,Kleinert:2002zn}.
However, in Brinkmann coordinates after the re-scaling \eqref{ps2}, 
$\det\,g=-1+{\cal O}(\lambda)$ 
and so to leading order in $R/m^2$ the determinant
factor is simply 1 and so plays no r\^ole. The same conclusion would
not be true in Rosen coordinates.

The $y^i$ fluctuations satisfy the eigenvalue equation 
\EQ{
\ddot y^i+\dot{\tilde u}^2
\sigma^2_iy^i-\frac{2\omega T\sigma_i}{m^2}\big(\tan(\sigma_iu_0+a_i)
\delta(\tau-\xi)+\tan(\sigma_iu_0-a_i)\delta(\tau)\big)y^i
=\lambda y^i-C\ ,
\label{eie}
}
where $C$ is the Lagrange multiplier that is determined by imposing the
constraint $\int_0^1d\tau\,y^i=0$. Now we see the utility of the Brinkmann
coordinates, because the equation \eqref{eie} is just that of a simple
harmonic oscillator and the non-trivial aspects of the problem lie
solely in the matching conditions at $\tau=0$ and $\tau=\xi$. On the
contrary, in Rosen coordinates the eigenvalue equation has
hypergeometric solutions and is not so straightforward to deal with. 
Consequently, we search for a solution in the form  
\EQ{
y^i(\tau)=\begin{cases}A_1\cos(\omega_1\tau)+B_1\sin(\omega_1\tau)-
C/\omega_1^2 & 0\leq \tau\leq
  \xi\\ A_2\cos(\omega_2\tau)+
B_2\sin(\omega_2\tau)-C/\omega_2^2 & \xi\leq\tau\leq 1\ .\end{cases}
\label{sop}
}
where $\omega_1^2=4\omega^2 T^2\sigma_i^2(1-\xi)^2/m^4-\lambda$ and 
$\omega_2^2=4\omega^2T^2\sigma_i^2
\xi^2/m^4-\lambda$. The matching conditions at $\tau=0(=1)$
and $\tau=\xi$ are the continuity of $y^i$ and the jumps
\SP{
\Delta\dot y^i(0)&=\frac{2\omega T\sigma_i}{m^2}
\tan(\sigma_iu_0-a_i)y^i(0)\ ,\\
\Delta\dot y^i(\xi)&=\frac{2\omega T\sigma_i}{m^2}
\tan(\sigma_iu_0+a_i)y^i(\xi)\ .
\label{fdd}
}
These conditions, along with
$\int_0^1d\tau\,y^i(\tau)=0$, determine the five unknowns
$A_i$, $B_i$ and $C$. A solution is only
possible if $\lambda$ satisfies a characteristic equation ${\cal F}
(\lambda)=0$. When ${\cal F}(\lambda)$ is suitably normalized, the
determinant of the fluctuation operator 
is given by ${\cal F}(0)$. This leads to the remarkably
simple formula for the determinant factor relative to flat
space:
\EQ{
{\cal Z}(\beta_l)=\prod_{i=1}^2
\sqrt{\frac{\beta_i^3\cos(\beta_i+a_i)\cos(\beta_i-a_i)}{\cos^2a_i\sin^3\beta_i
\cos\beta_i}}\ ,
\label{cuy}
}
where
\EQ{
\beta_i=\frac{\omega T\xi(1-\xi)\sigma_i}{m^2}\ .
}
Notice that ${\cal Z}\to1$ in the flat space limit $\sigma_i\to0$.

The remaining correlation function piece is determined by
the Green function
\EQ{
G_{ij}(\tau,\tau')=\big\langle y^i(\tau)y^j(\tau')\big\rangle\ .
}
It is clear that this is diagonal in the polarization indices, where
the diagonal components are the solution of the equation
\SP{
\Big[\partial_\tau^2+\dot{\tilde u}(\tau)^2\sigma^2_i-
&\frac{2\omega T\sigma_i}{m^2}\big(\tan(\sigma_iu_0+a_i)
\delta(\tau-\xi)\\ &+\tan(\sigma_iu_0-a_i)\delta(\tau)\big)\Big]G_{ii}
(\tau,\tau')
=-2i\delta(\tau-\tau')-C\ ,
}
We can find this
by a brute force solution similar to that above, imposing boundary
conditions so that $G_{ii}(\tau,\tau')$ is
continuous at $\tau=0$, $\tau'$ and $\xi$ 
and its derivative jumps by the appropriate amounts at $\tau=0$,
$\tau'$ and $\xi$. As before, $C$ is determined
by imposing $\int_0^1d\tau\,G_{ii}(\tau,\tau')=0$.
The solutions themselves are not very
illuminating and so we do not write them down here.
The Green function leads to the following remarkably simple formula
for\footnote{The limits $\tau\to\xi$ and $\tau'\to0$ have to be taken
  after the derivatives have been evaluated due to implicit dependence
  on $\xi$.}
\SP{
{\cal G}_{ij}(\beta_l)&=
\Big\langle \varepsilon_i[x(\xi)]\cdot\dot x(\xi)
e^{-i\omega\Theta[x(\xi)]}\,
\varepsilon_j[x(0)]\cdot
\dot x(0)e^{i\omega\Theta[x(0)]}
\Big\rangle\\
&=\frac{m^2\delta_{ij}}T\prod_{l=1}^2\sqrt{\frac{\cos^2a_l}
{\cos(\sigma_l\tilde u(\tau)+a_l)
\cos(\sigma_l\tilde u(\tau')+a_l)}}\\ &\qquad\times
\Big[\partial_\tau\partial_{\tau'}+\sigma_i
\tan(\sigma_i\tilde u(\tau)+a_i)\dot{\tilde u}(\tau)\partial_{\tau'}
+\sigma_i
\tan(\sigma_i\tilde u(\tau')+a_i)\dot{\tilde u}(\tau')\partial_{\tau}\\ & 
\qquad
+\sigma^2_i
\tan(\sigma_i\tilde u(\tau)+a_i)\tan(\sigma_i\tilde u(\tau')+a_i)
\dot{\tilde u}(\tau)
\dot{\tilde u}(\tau')\Big]G_{ii}(\tau,\tau')\Big|_{
\tau=\xi,\tau'=0}\\
&=\frac{2im^2\delta_{ij}}T\Big(
\delta(\xi)-\frac{\beta_i}{\sin\beta_i\cos\beta_i}\Big)
\prod_{l=1}^2\sqrt{\frac{\cos^2a_l}
{\cos(\beta_l+a_l)\cos(\beta_l-a_l)}}\ .
}
The $i$ here is crucial and appears because we are working in
Minkowski signature. 

Putting all these pieces together, the 
final result for the one-loop correction to the 
vacuum polarization is
\SP{
\Pi^\text{1-loop}_{ij}
&=\frac{\alpha}{4\pi}\int_0^\infty \frac{dT}{T}
ie^{-iT}\int_0^1d\xi\,\,{\cal Z}(\beta_l)
{\cal G}_{ij}(\beta_l)\\
&=\delta_{ij}\frac{\alpha m^2}{2\pi}\int_0^\infty \frac{dT}{T^2}
ie^{-iT}\int_0^1d\xi\,\left\{
1-\frac{\beta_i}{\sin\beta_i\cos\beta_i}
\prod_{l=1}^2
\sqrt{\frac{\beta_l^3}{\sin^3\beta_l\cos\beta_l}}\right\}\ .
}
The term $\delta(\xi){\cal Z}(\beta_l)/
(\cos\beta_1\cos\beta_2)$ has been replaced by $1$ since
\EQ{
\lim_{\xi\to0}{\cal Z}(\beta_i)\prod_{l=1}^2\sqrt{\frac{\cos^2a_l}
{\cos(\beta_l+a_l)\cos(\beta_l-a_l)}}=1\ .
}
It is remarkable that the result for the vacuum polarization 
is independent of $a_i$ so the
ambiguity in the choice of the null congruence has no effect on the
final result. It is especially noteworthy that the
divergences of the vertex operators due the 
singularities of the scalar amplitude at the caustics of the null
congruence are completely removed by quantum effects.

The mass-shell conditions for the two polarization states 
are modified by the one-loop correction to
\EQ{
\frac12(\omega^2-\vec k^2)+\Pi^\text{1-loop}_{ii}(\omega)=0\ .
}
The phase velocities are $v_\text{ph}=\omega/|\vec k|$ then and hence
the refractive indices for the two velocity eigenstates are
\SP{
&n_i(\omega)=\frac{|\vec
  k|}\omega=\frac{\sqrt{\omega^2+2\Pi^\text{1-loop}_{ii}(\omega)}}{\omega}
=1+\frac1{\omega^2}\Pi^\text{1-loop}_{ii}(\omega)+\cdots \\ &
=1+\frac{\alpha m^2}{2\pi\omega^2}\int_0^\infty \frac{dT}{T^2}
ie^{-iT}\int_0^1d\xi\,\left\{
1-\frac{\beta_i}{\sin\beta_i\cos\beta_i}
\prod_{l=1}^2
\sqrt{\frac{\beta_l^3}{\sin^3\beta_l\cos\beta_l}}
\right\}
\label{noi}
}
to order $\alpha$.
In particular, notice that the polarization vectors $\varepsilon_i$
correspond directly to the two velocity eigenstates. 

\section{Analysis and Interpretation}

The first remark is that the expression for the refractive indices
\eqref{noi} is completely
UV safe since the term in curly brackets behaves as $T^2$
for small $T$. This can be traced to the fact that we have
imposed the tree-level on-shell condition on the photon momentum. 

\begin{figure}[ht] 
\centerline{\includegraphics[width=3.5in]{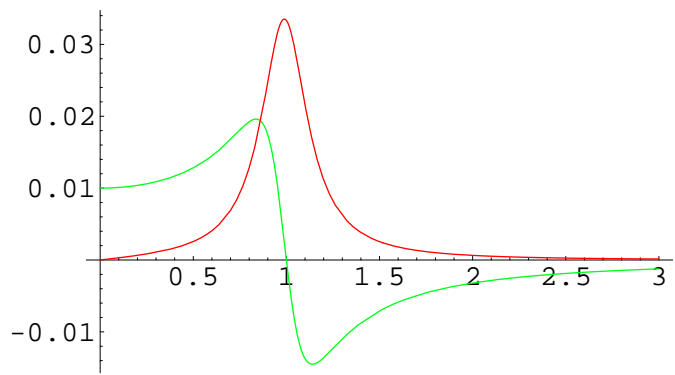}}
\caption{\footnotesize The real (green) and imaginary (red) parts of
  $n(\omega)-1$ for a simple model of a single absorption band with
  $\omega_0=1$, $\omega_p=0.1$ and $\gamma=0.3$.}\label{plot8}
\end{figure}

As we proceed, it is useful to have in mind the behaviour of the
refractive index in a simple model of a dissipative dielectric 
medium with a single absorption band.\footnote{This simple model forms
  the basis of many textbook discussions; for example, see
  Jackson \cite{jack}, chpt.~7.10.}  
This is modelled by an electric permittivity of the form
\EQ{
\epsilon(\omega)=1-\frac{\omega_p^2}{\omega^2-\omega_0^2+i\omega\gamma}\ .
}
where $\omega_0$ is the resonant frequency and $\gamma$ is the width.
For weak coupling,
\EQ{
n(\omega)=\sqrt{\epsilon(\omega)}=1-
\frac{\omega_p^2/2}{\omega^2-\omega_0^2+i\omega\gamma}+\cdots\ .
\label{jqq}
}
Written in the same form as \eqref{noi} as a $T$ integral, we have
\EQ{
n(\omega)=1-\frac{\omega_p^2}{\omega\omega_0}\int_0^\infty dT\,e^{-iT}\,
e^{-\omega\gamma T/(2\omega_0^2)}\sin(\omega T/\omega_0)\ .
\label{noa}
}
In this case, the $T$ integral is perfectly well defined without
the need for an $i\epsilon$ prescription. 
The real and imaginary parts of $n(\omega)-1$ 
are sketched in Figure~(\ref{plot8}).
 At low
frequencies the phase velocity is subluminal. At frequencies
$\omega\sim\omega_0$ the imaginary part of $n(\omega)$ has an
absorption peak and the phase velocity changes over to being
superluminal. At high frequencies, the phase velocity approaches 1 as
$1/\omega^2$. It is important to emphasize that the superluminal
phase velocity at high frequencies is not associated with a violation of
causality since asymptotically it approaches $c$.

\subsection{The analytic structure of the integrand}

\begin{figure}[ht] 
\centerline{\includegraphics[width=3in]{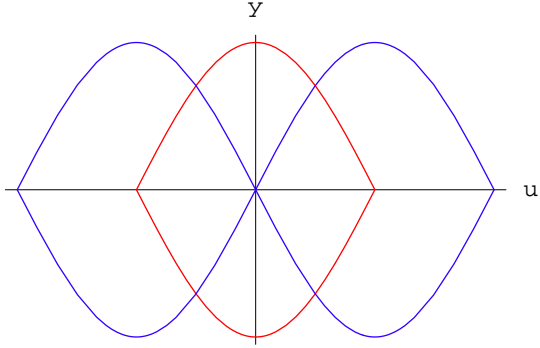}}
\caption{\footnotesize The $n=1$ (red) and $n=2$ (blue) zero modes for
  $\xi=\tfrac12$. The points $u=\pm u_0$ are conjugate points for 
$\gamma$.}\label{plot6}
\end{figure}

When we compare our result \eqref{noi} to the simple model of a
dissipative medium
\eqref{noa}, the most striking difference is the existence of
singularities in the integrand. When $\sigma_i$ is real, 
the integrand \eqref{noi}
has branch point singularities on the positive real axis at 
\EQ{
T=\frac{\pi m^2n}{2\xi(1-\xi)\sigma_i\omega}\ ,\qquad n=1,2,\ldots
\label{qs}
}
and the $T$ integral must be properly defined in
order to have a finite result. The correct procedure is to take the
contour to lie just below the real axis. It is significant
that these singularities arise from zeros of the
fluctuation determinant \eqref{cuy} and have a natural interpretation
in terms of zero modes, {\it viz\/}.~non-trivial solutions of \eqref{eie}
with zero eigenvalue $\lambda=0$. For the special case when
$\xi=\tfrac12$ these zero modes are particularly simple: $u=\tilde
u(\tau)$ as in \eqref{gsol} and
\EQ{
y^i(\tau)=\sin(2n\pi\tau)\ .
\label{ans}
}
The expression for $v(\tau)$ is then completely determined by solving
the geodesic equation
\EQ{
\ddot v+\sum_{i=1}^2\Big\{2\dot u\sigma_i^2y^i\dot y^i
+\frac{\omega
  T\sigma_i^2}{2m}y^{i2}\sec^2(\sigma_iu+a_i)\big(\delta(\tau-\xi)
-\delta(\tau)\big)\Big\}=0\ .
\label{veq}
}
These solutions are therefore associated with geodesics that are
arbitrarily close to $\gamma$ that intersect
$\gamma$ at both $u=\pm u_0$ and for $n>1$ at points in between. 
In other words, $u=\pm u_0$
are conjugate points on the geodesic $\gamma$. The $n=1$ and $n=2$
zero modes are illustrated in Figure~(\ref{plot6}).

\begin{figure}[ht] 
\centerline{\includegraphics[width=3in]{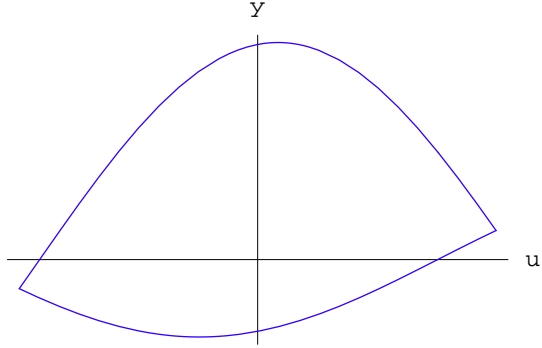}}
\caption{\footnotesize The $n=1$ zero mode for the case $\xi=\tfrac14$ and
  $a_i=0.2$.}\label{plot7}
\end{figure}

For generic $\xi$ the solutions are more complicated: again $u=\tilde
u(\tau)$ as in \eqref{gsol} while
\EQ{
y^i(\tau)=\begin{cases}A_1\sin(\sigma_i\tilde u(\tau))+B_1
\cos(\sigma_i\tilde u(\tau))& 0\leq\tau\leq \xi\\
A_2\sin(\sigma_i\tilde u(\tau))+B_2
\cos(\sigma_i\tilde u(\tau))& \xi\leq\tau\leq 1\ 
\end{cases}
\label{zm}
}
and $v(\tau)$ solves \eqref{veq}.
Imposing the continuity of $y^i$ and 
the conditions \eqref{fdd} implies
that for $n$ odd we must have 
\EQ{
A_1=A_2=\frac{1-2\xi}{1-\xi}B_1\tan a_i\ ,\qquad
B_2=-\frac{\xi}{1-\xi}B_1\ ,
}
and so there is only a single
zero mode. For $n$ even, however, there are two zero modes since there
are only two conditions on the four constants:
\EQ{
B_1=B_2\ ,\qquad
(1-\xi)A_1+\xi A_2=B_1\tan a_i\ .
}
Once again these solutions are formed from portions of two inequivalent geodesics which
intersect at $x^\mu(0)$ and $x^\mu(\xi)$: in other words, the points  
$x^\mu(0)$ and $x^\mu(\xi)$ are conjugate, however, for $a_i\neq0$ 
the conjugate points do not generally lie on $\gamma$. An example of
the first zero mode is illustrated in Figure~(\ref{plot7}).

Similarly the singularities on the imaginary $T$ axis that arise when
one of the $\sigma_i$ is imaginary can be understood in terms of these
zero modes, but with imaginary affine parameter $u\to iu$. It is
tempting to think that these solutions in 
imaginary affine parameter can be associated with
world-line instanton solutions. These kinds of instanton in the
world-line sigma model (not to be confused with instantons in the
original field theory) have been
discussed in the literature in the context of non-trivial
electromagnetic backgrounds and can be used to describe Schwinger pair
creation in that context \cite{Affleck:1981bm,Dunne:2005sx}. 
Here, we shall shortly see why this interpretation is
exactly right, since the singularities on the imaginary axis 
determine the imaginary
part of the refractive index which describes the dissipative nature of
the propagation that arises from the 
physical process $\gamma\to e^+e^-$. In the world-line instanton 
interpretation this is described as a tunneling process. 

\begin{figure}[ht] 
\centerline{\includegraphics[width=2.5in]{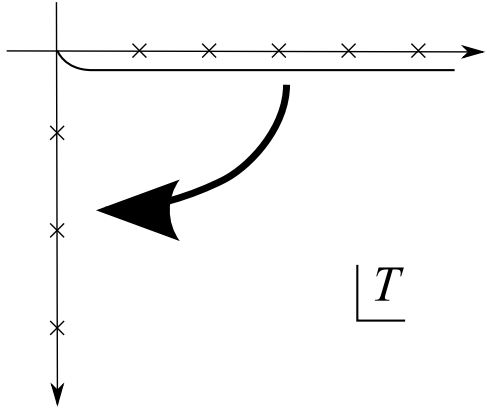}}
\caption{\footnotesize Wick rotating the integration contour to the
negative imaginary axis. The crosses represent branch point 
singularities.}\label{pic3}
\end{figure}

Now that we have explained the origin of 
the singularities, we can continue with the
analysis of \eqref{noi}. In order to produce a convergent integral, the
Wick rotation $T\to-iT$ can be performed as illustrated in 
Figure~(\ref{pic3}). 
Once Wick rotated, the resulting integral has the form of an
inverse Borel transform:
\EQ{
n_i(\omega)
=1-\frac{\alpha m^2}{2\pi\omega^2}\int_0^{\infty+i\epsilon}
\frac{dT}{T^2}
e^{-T}\int_0^1d\xi\,\left\{
1-\frac{\beta_i}{\sinh\beta_i\cosh\beta_i}
\prod_{l=1}^2
\sqrt{\frac{\beta_l^3}{\sinh^3\beta_l\cosh\beta_l}}
\right\}\ .
\label{noii}
}
When $\sigma_2$ is imaginary, 
{\it i.e.\/}~the Type II case, there are branch point
singularities on the integration contour and one has to be 
careful to take the contour to lie above the real axis as indicated by
the $i\epsilon$ prescription.

\begin{figure}[ht] 
\centerline{\includegraphics[width=2.5in]{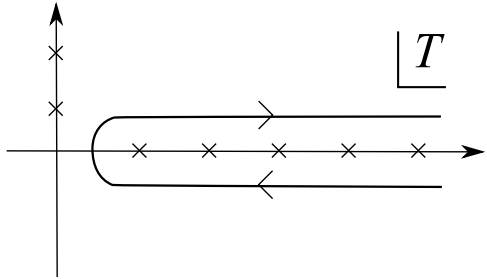}}
\caption{\footnotesize After Wick rotation, 
the contour $\EuScript C$ that computes the
  imaginary part of $n_i(\omega)$.}\label{pic4}
\end{figure}

Since the $T$ integral is of the
form $\int_0^{\infty+i\epsilon}dT\,e^{-T}f(T)$ where the function
$f(T)$ satisfies the reality condition
$f(T^*)^*=f(T)$, it follows that the imaginary part of the integral is
equal to $\tfrac12\int_{\EuScript C} dT\,e^{-T}f(T)$, where 
${\EuScript C}$ is a
contour that comes in from $\infty$ below the cut, goes round the first
branch point singularity and then goes to $\infty$ above the cut, as
illustrated in Figure~(\ref{pic4}).
Hence,
\EQ{
\text{Im}\,
n_i(\omega)
=-\frac{\alpha m^2}{4\pi\omega^2}\int_{\EuScript C} \frac{dT}{T^2}
e^{-T}\int_0^1d\xi\,\left\{
1-\frac{\beta_i}{\sinh\beta_i\cosh\beta_i}
\prod_{l=1}^2
\sqrt{\frac{\beta_l^3}{\sinh^3\beta_l\cosh\beta_l}}
\right\}\ .
\label{noe}
}

The imaginary part of the refractive index has an interesting
interpretation because it computes the probability
for pair creation, $\gamma\to e^+e^-$. In fact the 
total cross-section per unit volume, or 
inverse mean free path, is
\EQ{
\ell_\text{m.f.p.}^{-1}\thicksim\omega\,\text{Im}\,n(\omega)\ .
\label{mfp}
}
Notice that it is only
non-vanishing in the Type II case when there are singularities on the
real axis in the (Euclidean) $T$ plane. 
Earlier we pointed out that these
singularities correspond to non-trivial classical 
 loops with imaginary
affine parameter. We can now identify these solutions as world-line
instantons that describe the tunnelling process $\gamma\to
e^+e^-$. The fact that they occur only when $\sigma_2$
is imaginary is natural. Remember, when $\sigma_2$ is imaginary the
null congruence is defocusing in the $y^2$ direction. This suggests
the following intuitive picture:
after a virtual $e^+e^-$ pair is produced by
tunnelling, the pair then follow diverging geodesics and become real
particles. 

At low frequencies, the probability is dominated by the position of 
the first singularity at 
\EQ{
T=\frac{\pi m^2}{2\omega \xi(1-\xi)|\sigma_2|}\ ,
}
in Euclidean space, corresponding to the {\it fundamental\/} world-line
instanton which looks exactly like the red loop in 
Figure~(\ref{plot6}) in Euclidean time. 
The Euclidean action of a zero mode is simply $S_E=T$, and for small
$\omega$ we can use the steepest decent method to approximate the 
$\xi$ integral. The saddle-point is at $\xi=\tfrac12$
and hence the leading order behaviour
of \eqref{noe} for small $\omega$ will be of the form of an essential
singularity: 
\EQ{
\text{Im}\,n_i(\omega)\thicksim
\exp-\frac{2\pi m^2}{\omega|\sigma_2|}\ .
\label{fdr}
}

\subsection{The low frequency regime}

Low frequency means that $\omega^2 R/m^4\ll 1$. As a consequence
of this, the length of the loop \eqref{gsol}, $L\sim\omega/m^2$, is much
smaller than the curvature scale: 
$LR^{1/2}=\omega R^{1/2}/m^2\ll 1$. The leading order term 
in this limit will consequently be insensitive to the 
$u$ dependence of $h_{ij}(u)$ and so our result for this term is
valid for all background metrics and not just ones which yield symmetric plane
waves in the Penrose limit.

To calculate the expansion in $\omega$, we expand the Wick rotated 
integrand in \eqref{noii} in powers of $\omega$. The first term
in the expansion is $\omega$ independent:
\EQ{
n_i(\omega)=1-\frac{\alpha}{2\pi}
\frac{4\sigma_i^2+3\sum_{j=1}^2\sigma_j^2}{180m^2}+{\cal
  O}(\omega^2)
}
and so there is no dispersion in this limit.
Using \eqref{cc}, this can be written in terms of the curvature, and
the Newman-Penrose scalars, as 
\SP{
n_i(\omega)&=1-\frac{\alpha}{120\pi}\frac{R_{uu}}{m^2}
-\frac{\alpha}{90\pi}
\frac{R_{uiiu}}{m^2}+{\cal O}(\omega^2)\\
&=1-\frac{\alpha}{360\pi}{1\over m^2}\big(
10\Phi_{00}\mp 4|\Psi_0|\big)+{\cal O}(\omega^2)\ .
\label{huu}
}
for $i = 1,2$. In principle, this low frequency expression should follow 
from the terms in the low energy one-loop effective action of scalar QED
that are quadratic in the field strength $F$ and linear in the curvature. 
These terms have been calculated in spinor QED 
\cite{Drummond:1979pp,Shore:2002gw,Shore:2002gn} 
but not, to our knowledge, in scalar QED.\footnote{Even
given the effective action, one must be very careful in simplifying with 
integration by parts because the on-shell photon wavefunction does not 
fall off at infinity.}

It is interesting to consider the higher terms in the frequency
expansion in certain particular examples.
For example, for the case of a Type I conformally flat background,
$\sigma_1=\sigma_2=R^{1/2}$, the velocity eigenstates for both polarizations
have
\EQ{
n_i(\omega)=1-\frac{\alpha R}{2\pi m^2}\Big[\frac1{18}
-\frac{71}{14175}\frac{\omega^2R}{m^4}+
\frac{428}{189189}\Big(\frac{\omega^2R}{m^4}\Big)^2-
\frac{15688}{6891885}\Big(\frac{\omega^2R}{m^4}\Big)^3+\cdots\Big]\ .
\label{nbb}
}
This series is divergent but alternating and this is correlated
with the fact that it is Borel
summable, with the sum being defined by the 
convergent integral in \eqref{noii} which has no singularities 
on the real axis. Notice that $n_i(\omega)$ is real to all orders in the
expansion and since there are no cuts on the real 
axis the imaginary part vanishes,
as is evident in \eqref{noe}.

For the Type II Ricci flat background, $\sigma_1=i\sigma_2=R^{1/2}$,
one polarization is superluminal at low frequencies with
\EQ{
n_1(\omega)=1-\frac{\alpha R}{2\pi m^2}\Big[\frac1{45}
-\frac{37}{28350}\frac{\omega^2 R}{m^4}+
\frac{34}{85995}\Big(\frac{\omega^2R}{m^4}\Big)^2-
\frac{43}{135135}\Big(\frac{\omega^2R}{m^4}\Big)^3+\cdots\Big]\ .
\label{noo}
}
For the second, subluminal, polarization eigenstate,
\EQ{
n_2(\omega)=1+\frac{\alpha R}{2\pi m^2}\Big[\frac1{45}
+\frac{37}{28350}\frac{\omega^2R}{m^4}+
\frac{34}{85995}\Big(\frac{\omega^2R}{m^4}\Big)^2+
\frac{43}{135135}\Big(\frac{\omega^2R}{m^4}\Big)^3+\cdots\Big]\ .
\label{nop}
}
The first series \eqref{noo} is just the alternating version of
\eqref{nop}. In both cases the Borel
transforms have branch point singularities on the real axis and this is
indicative of an imaginary part \eqref{fdr} 
which vanishes to all orders in the $\omega^2R/m^4$ expansion. 

\subsection{The high frequency regime}

In the high-frequency limit $\omega^2R/m^4\gg1$, by re-scaling
 $T\to m^2T/(\omega \xi(1-\xi))$ and expanding
$\exp-m^2T/(\omega \xi(1-\xi))=1+\cdots$, 
we can show that the
$n_i(\omega)$ approach 1 like $1/\omega$:
\EQ{
n_i(\omega)=
1-\frac{\alpha C_i}{12\pi\omega}+{\cal O}\Big(\frac{\log\omega}{\omega^2}\Big)
\,
\label{srr}
}
where $C_i$ is the integral
\EQ{
C_i=\int_0^{\infty+i\epsilon} \frac{dT}{T^2}
\left\{
1-\frac{\sigma_iT}{\sinh\sigma_iT\cosh\sigma_iT}\prod_{l=1}^2
\sqrt{\frac{(\sigma_lT)^3}{\sinh^3\sigma_lT\cosh\sigma_lT}}
\right\}
\ .
\label{wee}
}
Notice that the behaviour of the subleading term is softer than $1/\omega^2$.
For the conformally flat case with $\sigma_1=\sigma_2\equiv R^{1/2}$, 
the integral \eqref{wee} can be evaluated exactly by contour 
integration yielding
\EQ{
C_i=\left(\frac13+\frac{7\pi^2}{36}\right)R^{1/2}\ ,
}
for both $i=1,2$.

For the Ricci flat Type II case (the vacuum gravitational wave), 
$\sigma_1=i\sigma_2\equiv R^{1/2}$, although we cannot evaluate $C_i$
analytically, there is an interesting relation
\EQ{
C_2=-iC_1^*\ ,
}
that follows from the definition of the integrals. A numerical
evaluation in this case gives
\EQ{
C_1=\big(0.22-0.014i\big)R^{1/2}\ ,\qquad
C_2=\big(0.014-0.22i\big)R^{1/2}\ ,
}
which implies that both polarization states are superluminal at high
frequencies. Hence, $n_2(\omega)$ must change from being greater than
$1$ to less than $1$ at some intermediate frequency. 

\subsection{Numerical analysis}

\begin{figure}[ht] 
\centerline{\includegraphics[width=2.5in]{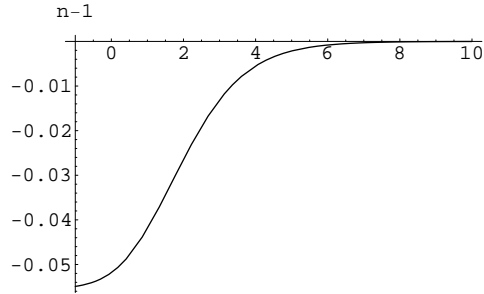}}
\caption{\footnotesize The behaviour of
$n_1(\omega)-1=n_2(\omega)-1$, in
  units of $\alpha R/(2\pi m^2)$, as a
  function of $\tfrac12\log\omega^2 R/m^4$ for the Type I conformally 
  flat case $\sigma_1=\sigma_2\equiv R^{1/2}$. The intercept 
$n_i(0)-1=-\tfrac1{18}\simeq-0.056$.}\label{plot1}
\end{figure}

\noindent{\it Type I:}~~~
In this case, the integrand \eqref{noii} 
is regular on the real axis and so the resulting refractive indices
are real and there is no pair creation. Figure~(\ref{plot1})~shows 
a numerical evaluation $n(\omega)$ for the conformally flat
background with $\sigma_1=\sigma_2\equiv R^{1/2}$.
\begin{figure}[ht]
\centerline{(a)\includegraphics[width=2.5in]{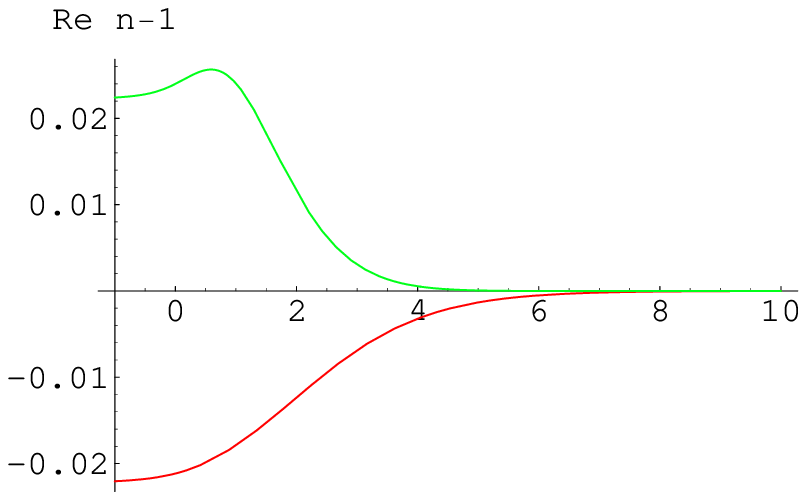}
\hspace{0.2cm}(b)\includegraphics[width=2.5in]{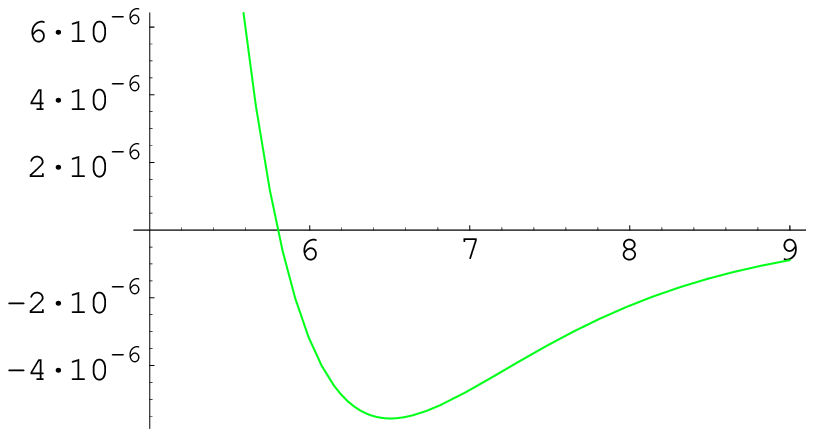}}
\caption{\footnotesize (a) The behaviour of
$\text{Re}\,n_i(\omega)-1$ ($i=1$ red, $i=2$ green), in
  units of $\alpha R/(2\pi m^2)$, as a
  function of $\tfrac12\log\omega^2 R/m^4$ for the Type II Ricci flat case
  (vacuum gravitational wave)
  $\sigma_1=i\sigma_2=R^{1/2}$. Notice that the intercepts 
$\text{Re}\,n_1(0)$ and $\text{Re}\,n_2(0)$ lie the equal amount
  $\tfrac1{45}\simeq0.023$
  below and above 1, respectively, in accordance with the polarization
sum rule\cite{Shore:1995fz}.  The resolution is not sufficient to 
show that the low-frequency subluminal photon 
becomes superluminal at high frequency. (b) A close-up of the region where 
$\text{Re}\,n_2(\omega)-1$ changes sign signalling that the 
subluminal photon becomes superluminal at sufficiently high 
frequency.}\label{plot2}
\end{figure}
\begin{figure}[ht] 
\centerline{(a)\includegraphics[width=2.5in]{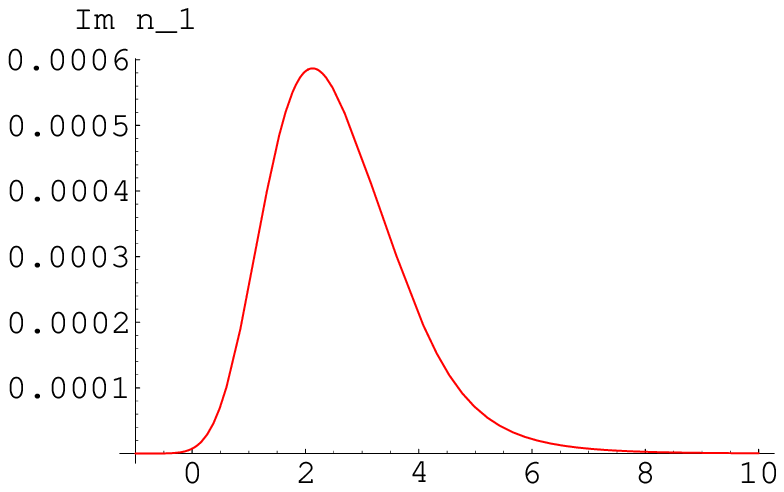}
\hspace{0.2cm}(b)\includegraphics[width=2.5in]{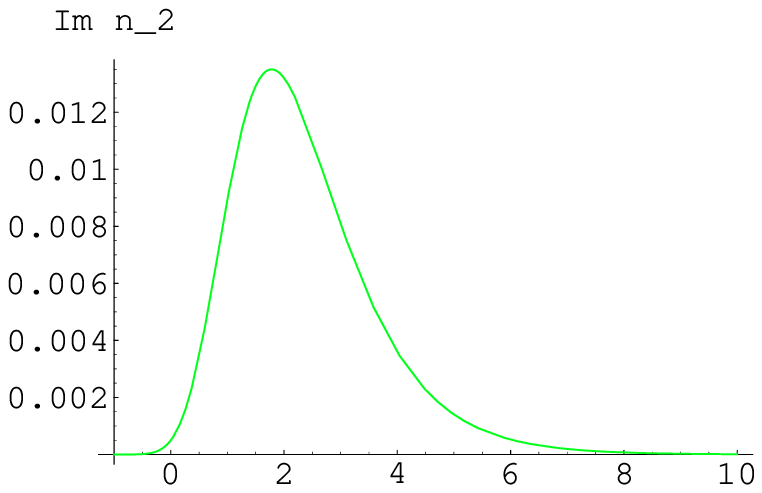}}
\caption{\footnotesize The behaviour of
(a) $\text{Im}\,n_1(\omega)$ and 
(b) $\text{Im}\,n_2(\omega)$ for the superluminal polarization state, in
  units of $\alpha R/(2\pi m^2)$, as a
  function of $\tfrac12\log\omega^2 R/m^4$ for the Type II Ricci flat 
case  (vacuum gravitational wave) $\sigma_1=i\sigma_2=R^{1/2}$. Notice that 
the subluminal polarization state, which is aligned with the defocusing
direction in the null congruence, has a much larger value of 
$\text{Im}\,n(\omega)$ than the superluminal state.}\label{plot3}
\end{figure}

\noindent{\it Type II:}~~~
In this case, the integrand \eqref{noii} has branch point
singularities on the real axis.
{}From the point of view of a numerical evaluation, it 
is therefore not useful to perform the Wick rotation. 
A useful alternative is to perform a ``half'' Wick rotation by 
rotating the contour of \eqref{noi} to lie along
$T\to(1-i)T/\sqrt2$. The resulting integral is convergent and 
can then be evaluated numerically. 
We find that the refractive
indices have both a real and imaginary part, as we anticipated earlier.
Figures~(\ref{plot2}) and (\ref{plot3}) 
show the real and imaginary parts of $n_i(\omega)$
for the example of a Ricci flat
background with $\sigma_1=i\sigma_2=R^{1/2}$. Notice that the
subluminal polarization state $n_2(\omega)$ behaves superficially
like our simple model of a dissipative medium in Figure~(\ref{plot8}). 

\section{Micro-Causality and the Kramers-Kronig Relation}

\begin{figure}[ht]
\centerline{(a)\includegraphics[width=2.5in]{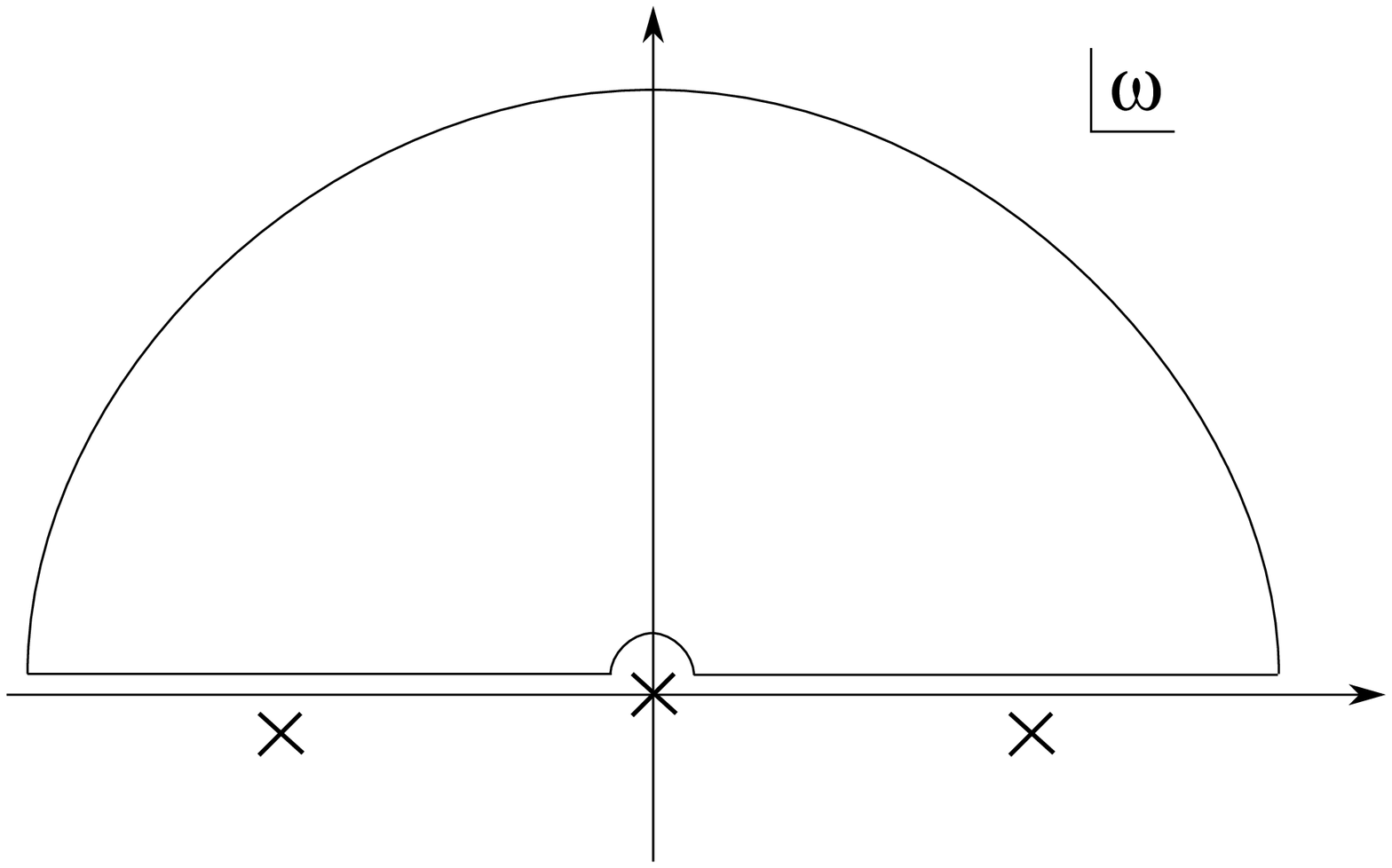}
\hspace{0.2cm}(b)\includegraphics[width=2.5in]{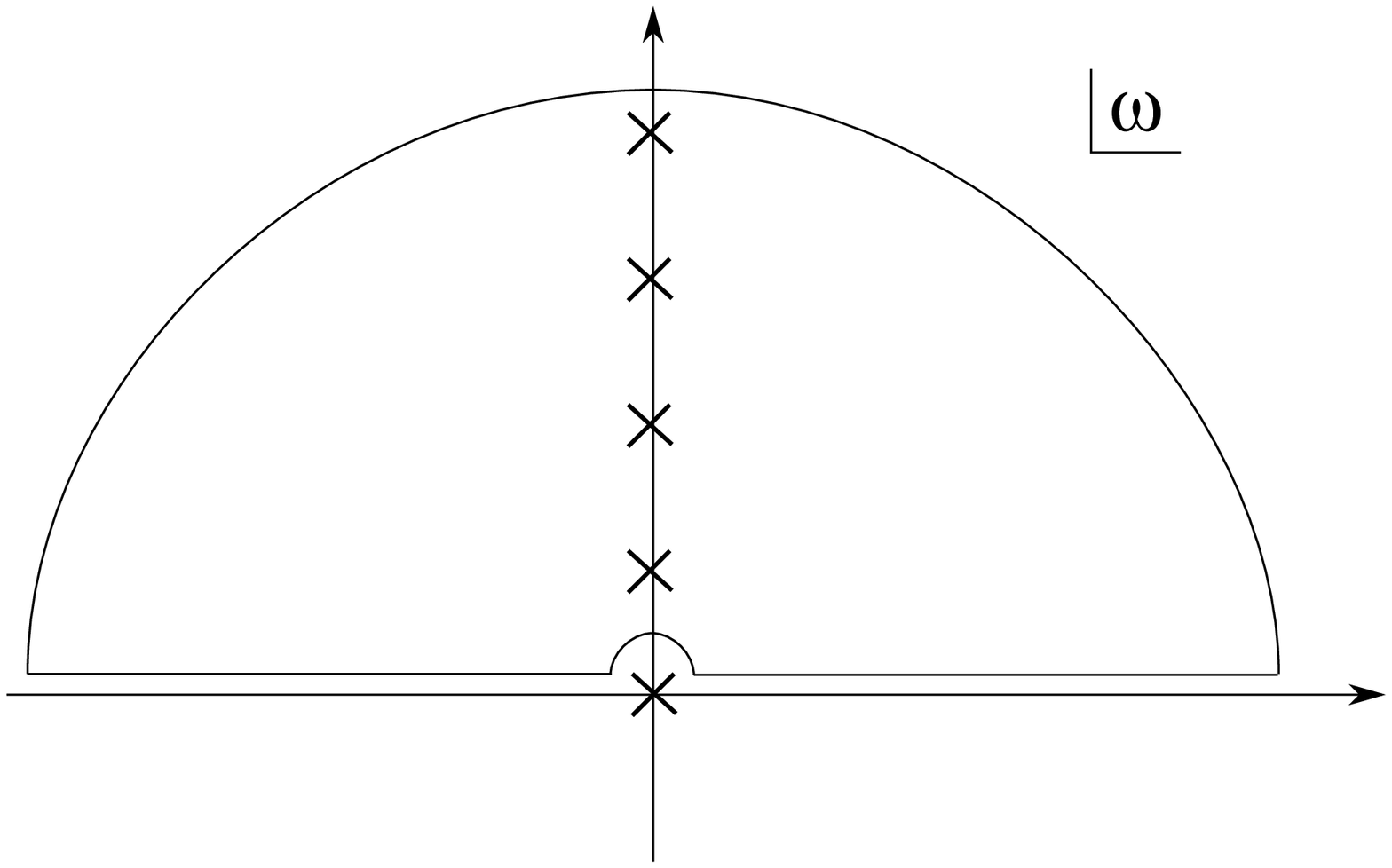}}
\caption{\footnotesize The integration contour for $\oint
  d\omega\,n(\omega)/\omega$ used in the
  proof of the KK relation for (a) the simple dissipative model with
  poles lying under the real axis (b)
  conformally flat case with poles on the imaginary axis.}
\label{pic15}
\end{figure}

Before we analyse our curved spacetime result, let us first consider
the simple model of a dissipative medium. In that case, from
\eqref{jqq} we see that $n(\omega)$
has simple poles in the lower-half plane at
$\omega=\pm\omega_0-i\gamma/2$ (for $\gamma\ll\omega_0$).
Hence, $n(\omega)$ is analytic in the upper-half
plane and the Kramers-Kronig relation is trivially satisfied. To see
this, consider $\int_{\cal C}
d\omega/\omega\,n(\omega)$ for a contour along the real axis, jumping
over the simple pole at $\omega=0$, completed by the large semi-circle
in the upper-half plane, illustrated in Figure~(\ref{pic15}). 
If $n(\omega)$ is analytic in
the upper-half plane, the total integral is 0 and so:
\SP{
0&=\int_\text{semi-circle}\frac{d\omega}\omega\,n(\omega)
-\pi i n(0)+{\cal P}\int_{-\infty}^\infty \frac{d\omega}\omega\,
n(\omega)\\
&=\pi i\big(n(\infty)-n(0)\big)+
{\cal P}\int_{-\infty}^\infty \frac{d\omega}\omega\,
n(\omega)\ .
\label{kk2}
}
Taking the imaginary part, and assuming that $n(\infty)$ and $n(0)$
are real and that $\text{Im}\,n(\omega)$ is an odd function,
immediately yields \eqref{kk}. It is a simple matter to check the
relation explicitly for the dissipative model \eqref{jqq}.

For curved spacetime this argument fails because there are
singularities on the imaginary axis which have to be included in
\eqref{kk2}, as illustrated in Figure.~(\ref{pic15}). 
For example, for the conformally flat Type I case,
$\sigma_1=\sigma_2=R^{1/2}$, the singularities are poles whose
residues must be included:\footnote{For more general examples, the
  singularities are branch points and the integration contour has to come
  around them from $\epsilon+i\infty$ down the imaginary axis
before going back to  $-\epsilon+i\infty$ and completing the semi-circle.} 
\EQ{
\pi i\big(n(\infty)-n(0)\big)+
{\cal P}\int_{-\infty}^\infty \frac{d\omega}\omega\,
n(\omega)=\text{pole contribution}\ .
\label{kk3}
}
Since in this case
$\text{Im}\,n(\omega)=0$, and including the contribution from the
poles on the imaginary axis, \eqref{kk3} becomes
\EQ{
\text{Re}\,n(0)-\text{Re}\,n(\infty)=\frac{\alpha R}{\pi m^2}
\int_0^\infty dT\,e^{-T}\,\int_0^1d\xi\,\big(\xi(1-\xi)\big)^2\,
\sum_{n=1}^\infty \text{Res}\,f(i\pi n/2)\ ,
}
where we have defined the function
\EQ{ 
f(x)=(1-x^4/(\sinh^4x\cosh^2x))/x^3\ .
}
The residue sum can be
regularized by considering $f(x)e^{ia x}$ and taking $a\to0$ at the
end. The result is
\EQ{
\text{Re}\,n(0)-\text{Re}\,n(\infty)=-\frac{\alpha R}{36\pi m^2}\ ,
}
which is in perfect agreement with \eqref{nbb} and \eqref{srr}. 
 Notice that we have established this result by interchanging the
order of the $\omega$ and $T$ integrals in which case the
singularities appear as poles on the imaginary axis. However, if we
perform the $T$ integral first, then the singularities become a branch
cut in $\omega$ from 0 to $\infty$ in the upper half plane.

The fact that $n(\omega)$ is not analytic in the upper-half plane is
intimately connected with the issue of micro-causality, as we now
explain. In our simple model of a dissipative 
medium, the Fourier transform of the susceptibility,
$\chi(\omega)=(\epsilon(\omega)-1)/(4\pi)$, 
\EQ{
G(t)=2\int_{-\infty}^\infty d\omega\,e^{-i\omega
  t}\chi(\omega)\ ,
}
plays the r\^ole of a response function: $\vec D(t)=\vec E(t)+\int
dt'\,G(t-t')\vec E(t')$. In the simple model, 
$n(\omega)$ and hence $\chi(\omega)$ is analytic in the 
upper-half plane and so, when $t<0$, we can compute the $\omega$ integral by
completing the contour with a semi-circle at infinity in the
upper-half plane. Since there are no singularities, the integral
vanishes implying $G(t)=0$: cause precedes
effect. Taking the explicit Fourier transform, we have 
\EQ{
G(t)=\frac{\omega_p^2}{\omega_0}e^{-\gamma t/2}\sin(\omega_0 t)
\theta(t)\
.
}
In the curved spacetime case, $n(\omega)$ is not analytic in the upper-half
plane and so it implies that 
the analogue of $G(t)$ will be non-vanishing for $t<0$. 

\begin{figure}[ht] 
\centerline{\includegraphics[width=2.2in]{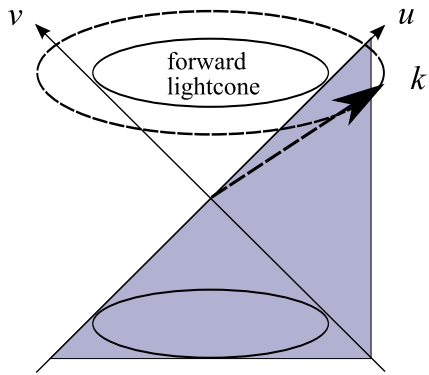}}
\caption{\footnotesize Including vacuum polarization effects, the
  photon momentum $k$ may lie outside the forward light cone ($u>0$,
  $v>0$) of its original null geodesic $v=0$. The potential violation of
  micro-causality implies that the retarded propagator is
  non-vanishing even for $v < 0$ (the shaded area), which
lies outside the forward light cone.}\label{pic14}
\end{figure}

We now place this simple analysis in the context of relativistic QFT,
where response functions are more properly understand in terms of
(retarded) propagators. The one-loop
vacuum polarization $\Pi^\text{1-loop}$ contributes to the propagator via 
$\Delta=\Delta^\text{tree}-\Delta^\text{tree}
\Pi^\text{1-loop}\Delta^\text{tree}
+\cdots$. In a real space picture, the issue of micro-causality rests on the
fact that the retarded propagator $\Delta_\text{ret}(x)$ is only
non-vanishing in, or on, the forward light cone.\footnote{When we talk
in the following about the ``light cone'' we mean the geometrical
null surface defined by the metric $g_{\mu\nu}$.}
In prosaic language, an external source can only influence the fields 
in the future. For instance, in the present context
the real space tree-level retarded propagator
$\Delta_\text{ret}^\text{tree}$ is only non-vanishing on the forward
light cone. However, what about the one-loop correction? 
Notice that we have only calculated 
$\Pi^\text{1-loop}(\omega)$ on-shell in momentum
space and this means that we do not have access to 
the complete one-loop real space propagator. However, we can 
perform
the Fourier transform with respect to $\omega$, which determines the 
propagator as a function of the null coordinate $v$:
\EQ{
\Pi^\text{1-loop}(v)=
\int_{-\infty}^{\infty} d\omega\,e^{-i\omega v}\Pi^\text{1-loop}(\omega)
\ .
}
This is a retarded quantity if the integration contour is taken to 
avoid singularities by veering into the upper-half plane,
when $v<0$, and the lower half plane, when $v>0$. 
For QFT in flat spacetime,
$\Pi^\text{1-loop}(\omega)$ is analytic in the upper-half plane and so
when $v<0$ one computes the $\omega$ integral by completing the contour with a
semi-circle at infinity in the upper-half plane. Since there are no
singularities in the upper-half plane, the integral
vanishes and consequently $\Pi^\text{1-loop}_\text{ret}(v)=0$ for
$v<0$. This is consistent with the fact that
the region $v<0$ lies outside the forward light
cone. Hence, in this case micro-causality is preserved as a
consequence of analyticity in the upper-half plane in frequency space.
In curved spacetime, on the
contrary, $\Pi^\text{1-loop}(\omega)$ is not
analytic in the upper-half plane and consequently it would seem that 
the one-loop retarded propagator $\Pi^\text{1-loop}_\text{ret}(v)$ must
receive contributions from the region $v<0$ which lies 
outside the forward light cone. See Figure~(\ref{pic14}).

The idea
here, is that $\omega$ is identified with one of the lightcone momenta
$p_+$. When the photon is on-shell at tree level, 
the other component vanishes, $p_-=0$. Now by giving
$p_-$ a small positive imaginary part a non-analyticity in the
upper-half $p_+$ plane would be in the region 
$\text{Im}\,p_+/\text{Im}\,p_->0$ which means that the retarded
propagator is non-vanishing outside the lightcone. However the
loophole in this reasoning is that once the $T$ and $\xi$ integrals
have been performed, the non-analyticities of $n(\omega)$ as a function
of $\omega=p_+$ arises as branch
cuts joining $p_+=0$ to $p_+=i\infty$. When taken off-shell, it may be
that the branch cut slips into the causally safe region 
$\text{Im}\,p_+/\text{Im}\,p_-<0$. The only way to really settle this issue
is to perform a calculation of the vacuum polarization with the
momentum {\it off-shell\/} 
\cite{Dubovsky:2007ac}.\footnote{{\it Note added}: We have now 
completed a full off-shell calculation of the vacuum polarization tensor
and find that precisely this behaviour occurs. The branch point at
$p_+=i\infty$ is shifted to $p_+=m^2/p_-$ with
$\text{Im}\,p_+/\text{Im}\,p_-<0$. Full details will be presented
elsewhere \cite{usagain}.}

One final point to emphasize is that in the Type II examples 
where the imaginary part of the
refractive index is non-vanishing, it is positive as expected from the
optical theorem\footnote{It is amusing to note that
here we are applying the optical theorem
in its original context of the refractive index from which the name of
the theorem is derived.} 
which relates it to the total cross-section per unit volume, or 
inverse mean free path as in \eqref{mfp}.
In this sense, spacetime acts as an ordinary dissipative
medium. However, unlike the simple dissipative model where the imaginary part
of the refractive index falls off as $1/\omega^3$, for curved spacetime the 
imaginary part falls off as $\alpha R^{1/2}/\omega$. This implies that the 
mean free path saturates to a constant $\sim 1/(\alpha R^{1/2})$.

\section{Remarks on General Backgrounds}

In this section, we consider some of the features of our analysis which
apply in more general non-symmetric plane wave backgrounds when
$h_{ij}(u)=R_{u iju}(u)$ 
does not take the special form \eqref{ca}. For the moment, we shall
assume that the Penrose limit is non-trivial, {\it i.e.\/}~not flat
space. Our purpose is to highlight how the resulting 
behaviour of the refractive index depends to a large extent on the 
properties of the null congruence.

At the beginning of Section 5, we found that the analytic structure
of the integrand \eqref{noi} -- more precisely the positions of the
singularities -- could be traced to the existence of zero modes of the
$y^i$ fluctuation equations. In the general case, 
the equation for these zero modes is
\EQ{
\ddot y^i+\dot{\tilde u}^2h_{ij}y^j+\frac{2\omega T}{m^2}
\Omega_{ij}y^j\delta(\tau-\xi)
-\frac{2\omega T}{m^2}\Omega_{ij}y^j\delta(\tau)=-C\ .
}
Considering the solutions of these 2nd order equations in the two regions 
$0\leq\tau\leq\xi$ and $\xi\leq\tau\leq1$, and including $C$ and $T$, there 
are generically nine unknowns to be fixed, up to 
overall scaling of the solution.
At $\tau=0$ and $\tau=\xi$ there are a total of eight boundary conditions on 
$y^i$ and $\dot y^i$, with the constraint $\int_0^1d\tau\,y^i=0$
providing a ninth condition. In general, we therefore expect
solutions to exist only for particular values of $T$. This is
exactly what we found for the symmetric case where the special values
of $T$ are given in \eqref{qs}.
Although, for $n$ even we found two zero modes rather than
the one expected.

As we have seen, the existence of the zero modes is related to
the behaviour of the null congruence. This can be made more concrete
by looking at the special case when
$\xi=\tfrac12$ (which is, in any case, picked out by the saddle-point
method described above). In this case, by symmetry we expect a solution of
the form that we found in the symmetric plane wave case, \eqref{ans},
but where the transverse geodesic deviation vector $y^i(u)$ satisfies the 
equation for a {\it Jacobi field\/} along the geodesic $\gamma$:
\EQ{
\frac{d^2y^i}{du^2}=-h_{ij}y^j\ ,
\label{jac}
}
subject to the boundary conditions $y^i(\pm u_0)=0$. Because of the
latter, the source terms vanish and so the derivatives $dy^i/d\tau$ must be 
continuous at $\tau=0(\equiv 1)$ and $\tau=\xi$, a fact that follows directly 
from the ansatz \eqref{ans}. In addition, the constraint 
$\int_0^1d\tau\,y^i=0$ is automatically satisfied and the Lagrange
multiplier $C$ vanishes. The classical solution is then $u=\tilde
u(\tau)$, (with a slight
abuse of notation) $y^i(\tau)=y^i(\tilde u(\tau))$ and $v(\tau)$
satisfies its own geodesic equation. Since
\eqref{jac} is a second order linear equation there will in general be solutions only for particular values of $u_0=\omega T/(4m^2)$.
Since $y^i(u)$ vanishes at $u=\pm u_0$, at least when the special values of 
$T$ are real, these points are precisely {\it conjugate points\/} along 
$\gamma$. Hence, the existence of zero modes (for real values of $T$) 
is tied directly to the existence of conjugate points. Notice, however, that
the special values of $T$ for which zero modes exist are not necessarily
real. This is exactly what happens in the Type II examples, where the
singularities corresponding to world-line instantons have imaginary
$T$. 

The zero modes dictate the analytic structure of 
the $T$ integral which in turn determines the
nature of the physics. In particular, the singularities along the
real $T$ axis play a prominent r\^ole because, as we have seen, they
are responsible for the non-trivial analytic structure of 
$n(\omega)$. Moreover, as we have argued above, 
zero modes for real $T$ correspond directly to the existence of conjugate
points along $\gamma$. But the existence of these conjugate points, as
explained in Section 3, is generic. Therefore we are led to the
following conclusion:
\begin{quote}
{\bf Conclusion:} violations of analyticity and the Kramers-Kronig
relation are generic and can be traced to the focusing nature of 
null geodesics and the existence of conjugate points implied by the
null energy condition. 
\end{quote}

As already mentioned, the Penrose limit is ideally suited to the analysis 
of photon propagation in arbitrary background spacetimes. Many of the
characteristic features of superluminal low-frequency propagation previously
found in specific examples, including Schwarzschild, Reissner-Nordstr\"om
and Kerr black holes \cite{Drummond:1979pp,Daniels:1993yi,Shore:1995fz} 
as well as gravitational waves \cite{Drummond:1979pp,Shore:2000bs},
can be seen directly in the Penrose limit.
For example, a maximally symmetric spacetime such as de Sitter has 
vanishing $\Phi_{00}$ and $\Psi_0$ and the low-frequency phase velocity
$v_{\rm ph}(0)$ receives no correction from vacuum polarization.
Using our formalism, we see immediately that at leading order in $R/m^2$
this result holds for {\it all\/} frequencies since the Penrose limit of a
maximally symmetric spacetime is flat \cite{Blau2}.

In Schwarzschild spacetime, we have previously found that while a photon
following a general null geodesic may experience a superluminal shift
in $v_{\rm ph}(0)$, the effect vanishes for purely radial geodesics. 
(In fact, this remains true for photons following principal null geodesics
\cite{Chandra} for any Petrov type D spacetime such as Schwarzschild or Kerr,
again due to the vanishing of the corresponding $\Phi_{00}$ and $\Psi_0$.) 
This is clear in our formalism. The Penrose plane wave limit 
for the Schwarzschild metric is, in Brinkmann coordinates,
\EQ{
ds^2 =2du dv + {3m L^2\over r(u)^5}\bigl((y^1)^2 - (y^2)^2\bigr)du^2
- (dy^1)^2 - (dy^2)^2,
\label{schwarz}
} 
where $L$ specifies the angular momentum and $r(u)$ is given by the solution 
of the geodesic equation. We see immediately that for radial trajectories
the Penrose limit is flat and so, at least at ${\cal O}(R/m^2)$, the 
phase velocity $v_{\rm ph}(\w)$ remains equal to $c$ for all frequencies,
not just in the low-frequency limit. Clearly, in such cases where
the Penrose limit is flat, the expansion \eqref{penexp} gives a systematic
way to go beyond leading order in $R/m^2$. An interesting feature is the
existence of a ``peeling theorem''\cite{Blau2}, whereby successive
orders in the Penrose expansion involve the curvatures $\Psi_0, \Psi_1,
\ldots \Psi_4$.

This gives a first glance at the power of the Penrose plane wave geometry
combined with the world-line sigma model approach. Moreover, other general 
features of null congruences will play an important r\^ole. For example, 
we have been implicitly assuming that the geodesics are complete so that the
affine parameter varies from $-\infty$ to $+\infty$. However, there are 
spacetimes where certain null geodesics are incomplete and the affine
parameter has a finite limiting value. This usually signals the existence
of a spacetime singularity, as for example in the case of Schwarzschild orbits
for $L$ less than a critical value, where the Penrose limit becomes singular
\cite{Blau2}. Clearly, this can affect the zero modes in the sigma model
and therefore the singularities and asymptotic behaviour of the refractive
index. The r\^ole of horizons in relation to the Penrose limit also
deserves investigation. All of these issues will be considered in detail
elsewhere.

\section{Conclusions}

In this paper, we have for the first time evaluated the
non-perturbative frequency
dependence of the vacuum polarization for QED in curved spacetime and 
determined the corresponding refractive index for photon propagation.
In so doing, we have resolved the outstanding problem in ``quantum 
gravitational optics''
\cite{Shore:2003zc,Shore:2007um}, {\it viz\/}.~how to reconcile
the prediction of a superluminal phase velocity at low frequency with 
causality. Remarkably, the resolution involves the violation of 
analyticity calling into question micro-causality in curved spacetime.

These results have been achieved by combining two powerful techniques:~
(i)~the world-line sigma model, which enables the non-perturbative
frequency dependence of the vacuum polarization to be evaluated by a
saddle-point expansion around a geometrically motivated classical solution,
and (ii)~the Penrose plane wave limit, which encodes the relevant tidal 
effects of spacetime in the neighbourhood of the original null geodesic
traced by the photon.

The form of the refractive index reflects the nature of the background 
spacetime. We identify two classes. In Type I backgrounds, which include
conformally flat spacetimes,\footnote{Note that the Penrose limit 
of a conformally flat spacetime is also conformally flat. Similarly for 
Ricci flat, and also locally symmetric, spacetimes \cite{Blau2}.}
both photon polarizations are superluminal at low frequencies, but the 
phase velocity approaches $c$ at high frequency. The imaginary part of the
refractive index vanishes. In Type II backgrounds, which include
Ricci flat spacetimes, photon propagation may display birefringence
with one superluminal and one subluminal polarization at low frequency.
In both cases, however, the high frequency phase velocity is $c$.
The refractive index develops an imaginary part, indicating a non-zero
probability for pair creation, $\c \rightarrow e^+ e^-$.
Since the high-frequency limit of the phase velocity is identified with
the wavefront velocity $v_{\rm wf}$, which is the ``speed of light'' relevant 
for causality, we see explicitly how superluminal propagation in the
low-frequency theory is compatible with causality.

Although these results were obtained using the Penrose limit in locally
symmetric spacetimes, they are expected to be generally true. The reason is
that the analytic properties of the refractive index can be related in the
world-line sigma model formalism to general results in the theory of 
null congruences. In particular, the distinction between Type I and Type II
spacetimes is whether the null geodesics in the congruence focus in both
transverse directions (Type I), or focus in one and defocus in the other 
(Type II). The result that at least one direction is focusing is a 
consequence of the null energy condition. The presence of a focusing 
direction in the congruence then implies the existence of conjugate
points, which leads to the existence of zero modes and ultimately yields
poles in the
refractive index in the upper-half complex plane, violating the analyticity
assumptions used to derive the Kramers-Kronig dispersion relation.
The violation of this dispersion relation in turn allows $n(\infty) > n(0)$ 
and removes the apparent paradox of having a superluminal phase velocity
$v_{\rm ph}(0) > c$ while the wavefront velocity $v_{\rm wf} =
v_{\rm ph}(\infty) = c$.

This is potentially the most far-reaching conclusion of this paper.
The null energy condition and the general relativistic theory of null
congruences necessarily imply a non-analyticity of the refractive
index, although the full implications of this for micro-causality and
the other axioms of S-matrix theory will only follow from an 
off-shell extension of the calculation.

The loss of analyticity in $n(\w)$, or more generally in forward scattering
amplitudes, also has important implications for the idea that constraints
may be placed on the parameters of a low-energy effective field theory by
the requirement that it admits a consistent UV completion
\cite{Shore:2007um,Adams:2006sv,Distler:2006if}. These constraints
are typically derived either by requiring the absence of superluminal
effects in the low-energy theory or assuming analyticity in dispersion
relations involving forward scattering amplitudes. While these remain valid in 
flat spacetime, we have shown that they are not applicable to fundamental
UV theories involving gravity, including string theory.

The full implications of the calculation of the refractive index and
the issues of causality and micro-causality
remain to be explored, 
especially in relation to horizons and singularities. 
The significance of the UV-IR mixing 
whereby the high-frequency limit probes the global properties of the null 
geodesic congruence also deserves to be better understood. 
What is clear, however, is that the results described here will have a 
significant impact on our understanding of quantum field theories
involving gravity.

\vskip0.5cm

We would like to thank Asad Naqvi for many useful conversations and 
Sergei Dubovsky, Alberto Nicolis, Enrico Trincherini and 
Giovanni Villadoro for pointing out the necessity of working
off-shell in order to completely settle the question of micro-causality.
TJH would also like to thank Massimo Porrati for a helpful discussions
and Fiorenzo Bastianelli for explaining some details 
of his work on the world-line formalism. This work was supported in part
by PPARC grant PP/D507407/1.

\startappendix

\Appendix{Power Counting}

In this Appendix, we prove in an alternative way 
one of the key results of this paper: that
each loop in the world-line QFT comes with a power of $R/m^2$ and so
loops are suppressed in the limit of weak curvature $R\ll m^2$.
In order to assess the behaviour of a given graph in perturbation
theory, it is useful to re-scale $T\to T/m^2$ and then
 $\tau\to \tau T$ and $x\to \sqrt T x$
so that the world-line action can be split as
\EQ{
S=\frac14\int_0^1d\tau\,\eta_{\mu\nu}\dot x^\mu\dot
x^\nu+S_\text{pert}\ ,
}
where a typical term in 
$S_\text{pert}$ arises from expanding the metric around flat space at
the point $x_0=0$; schematically,
\EQ{
\int_0^1d\tau\,\left(\frac{R}{m^2}\right)^{n/2}x^n\dot x^2\ ,
\label{ioo}
}
where $R^{n/2}$ denotes powers of the Riemann tensor and its
derivatives of mass dimensions $n$. The vertex
behaves as $(R/m^2)^{n/2}$ and has
$n+2$ legs. In addition, we have the
exponential factors $\omega\Theta$ which we can view as additional
vertices of the form
\EQ{
\omega\Theta\thicksim \frac\omega m
\sum_n\left(\frac R{m^2}\right)^{n/2}x^{n+1}\ .
\label{frr}
}

Consider such a graph with $E$ external legs, $I$ internal
legs and $V$ vertices. If the graphs consists of $N_{n}$ vertices of
the form \eqref{ioo} and $S_n$ vertices of the form \eqref{frr}, then
\EQ{
\sum_{n}\big((n+2)N_{n}+(n+1)S_n\big)=2I+E\ ,\qquad V=
\sum_{n}\big(N_{n}+S_n\big)\ .
}
The graph behaves as 
\EQ{
\omega^{\sum_nS_n}m^{-\sum_n(nN_n+(n+1)S_n)}
R^{\sum_{n}n(N_n+S_n)/2}
=\left(\frac{\omega^2 R}{m^4}\right)^{\sum_nS_n/2}
\left(\frac R{m^2}\right)^{I-V+E/2}\ .
}
Now we use the topological identity, $L=I-V+1$, where $L$ is the
number of loops, to equate this to
\EQ{
\left(\frac{\omega^2R}{m^4}\right)^{\sum_nS_n/2} \left(\frac R{m^2}
\right)^{L-1+E/2}\ .
}
So each loop brings a factor of $R/m^2$. For example, 
the partition function ${\cal Z}$ has
$E=0$ and since the tree-level contribution is the classical action for the
saddle point which vanishes, the leading order term comes
from one loop and is an arbitrary function of $\omega^2 R/m^4$. The
expansion around the classical saddle-point solution sums up all the
one-loop graphs with arbitrary $\omega$ insertions.
The leading order contribution to the Green's function piece, which has
$E=2$, comes from tree level. Once again, the
expansion around the classical saddle-point solution sums up all these
tree graphs with arbitrary $\omega$ insertions.

\end{document}